\documentclass[format=sigconf, nonacm, screen=True]{acmart}
\usepackage{tikz}
\usepackage{subcaption}
\usepackage{threeparttable}
\usepackage{array}
\usepackage{geometry}
\usepackage[utf8]{inputenc}
\usepackage{graphicx} 
\usepackage{pgfplots}
\usepackage{caption}
\usepackage{graphicx}
\usepackage{bbm}
\usepackage{threeparttable}
\usepackage{soul}
\usepackage[misc]{ifsym}
\usepackage[capitalize]{cleveref}
\usepackage{rotating}
\usepackage{float}
\usepackage{longtable}
\usepackage{array}
\usepackage{amsmath}
\usepackage{tabularx}
\usepackage{ragged2e}
\usepackage{lipsum}
\usepackage{longtable}
\usepackage{enumitem}
\usepackage{setspace}
\usepackage{xcolor} 
\usepackage{balance}
\pgfplotsset{compat=1.17}
\usepackage[subtle,tracking=normal, wordspacing=normal,mathdisplays=tight]{savetrees}
\usepackage{tipa}

\newif\ifshowcomment
\showcommenttrue
\ifshowcomment
\newcommand{\luyao}[1]{\textcolor{blue}{[luyao] #1}}

\else
\newcommand{\luyao}[1]{}
\fi

\begin{document}

\title[Understand Waiting Time in Transaction Fee Mechanism:  An Interdisciplinary Perspective]{Understand Waiting Time in Transaction Fee Mechanism: \\ An Interdisciplinary Perspective}

\author[Luyao Zhang*]{Luyao Zhang}
\affiliation{%
  \department{Data Science Research Center and Social Science Division}
  \institution{Duke Kunshan University}
  \country{China}
}

\authornote{Joint first and corresponding authors: \newline
Fan Zhang (email: f.zhang@yale.edu, address: 51 Prospect St, New Haven, CT 06520, United States) and Luyao Zhang (email: lz183@duke.edu, address: Duke Kunshan University, No.8 Duke Ave. Kunshan, Jiangsu 215316, China.) }

\authornote{Luyao Zhang completed the research during university vacation time as an independent project and is also with SciEcon CIC, 71-75 Shelton Street, Covent Garden, London, United Kingdom, WC2H 9JQ}

\author[Fan Zhang*]{Fan Zhang}
\affiliation{%
  \department{Department of Computer Science}
  \institution{Yale University}
  \country{United States}
}
\authornotemark[1]

\begin{abstract}
\begin{quote}
I would rather discover one causal law than be the King of Persia.\\---Democritus, 460---378 B.C.~\cite{pearl2002reasoning} 
\end{quote}
Blockchain enables peer-to-peer transactions in cyberspace without a trusted third party. The rapid growth of Ethereum and smart contract blockchains generally calls for well-designed Transaction Fee Mechanisms (TFMs) to allocate limited storage and computation resources. However, existing research on TFMs must consider the waiting time for transactions, which is essential for computer security and economic efficiency. Integrating data from the Ethereum blockchain and memory pool (mempool), we explore how two types of events affect transaction latency. First, we apply regression discontinuity design (RDD) to study the causal inference of the Merge, the most recent significant upgrade of Ethereum. Our results show that the Merge significantly reduces the long waiting time, network loads, and market congestion. In addition, we verify our results' robustness by inspecting other compounding factors, such as censorship and unobserved delays of transactions via private changes. Second, examining three major protocol changes during the merge, we identify block interval shortening as the most plausible cause for our empirical results. Furthermore, in a mathematical model, we show block interval as a unique mechanism design choice for EIP1559 TFM to achieve better security and efficiency, generally applicable to the market congestion caused by demand surges. Finally, we apply time series analysis to research the interaction of Non-Fungible token (NFT) drops and market congestion using Facebook Prophet, an open-source algorithm for generating time-series models. Our study identified NFT drops as a unique source of market congestion--holiday effects---beyond trend and season effects. Finally, we envision three future research directions of TFM. Our study contributes to the interdisciplinary research at the intersection of computing and economic science around the application of blockchain technology, including computer security, distributed systems, market design, causal inferences, and time series analysis. Our findings shed light on a new direction of TFM that improves blockchain security and efficiency through a hybrid method of empirical and theoretical study. 
\end{abstract}




\maketitle

\section{Introduction}
Blockchain enables peer-to-peer transactions in cyberspace without a trusted third party~\cite{nakamoto2008bitcoin, casey2018blockchain}.
The rapid growth of Ethereum~\cite{buterin_2013_ethereum} and smart contract blockchain in general~\cite{vujivcic2018blockchain,mohanta2018overview} calls for well-designed Transaction Fee Mechanisms (TFMs)~\cite{Vitalik2019, roughgarden_2020_transaction,Roughgarden2021, ferreira_2021_dynamic,chung_2022_foundations,easley_2019_from, harvey_2021_defi} to allocate limited storage and computation resources~\cite{buterin_2018_blockchain}. \citet{ante2023time} document the important role of transaction fees in affecting the economic activities of various subsystems---including bridges, centralized exchange (CEX), decentralized exchange (DEX), maximum extractable value (MEV), non-fungible tokens (NFTs), and  stablecoins---on Ethereum blockchain network using time-series analysis of Granger causality. 

\begin{figure}
	\includegraphics[width = \linewidth]{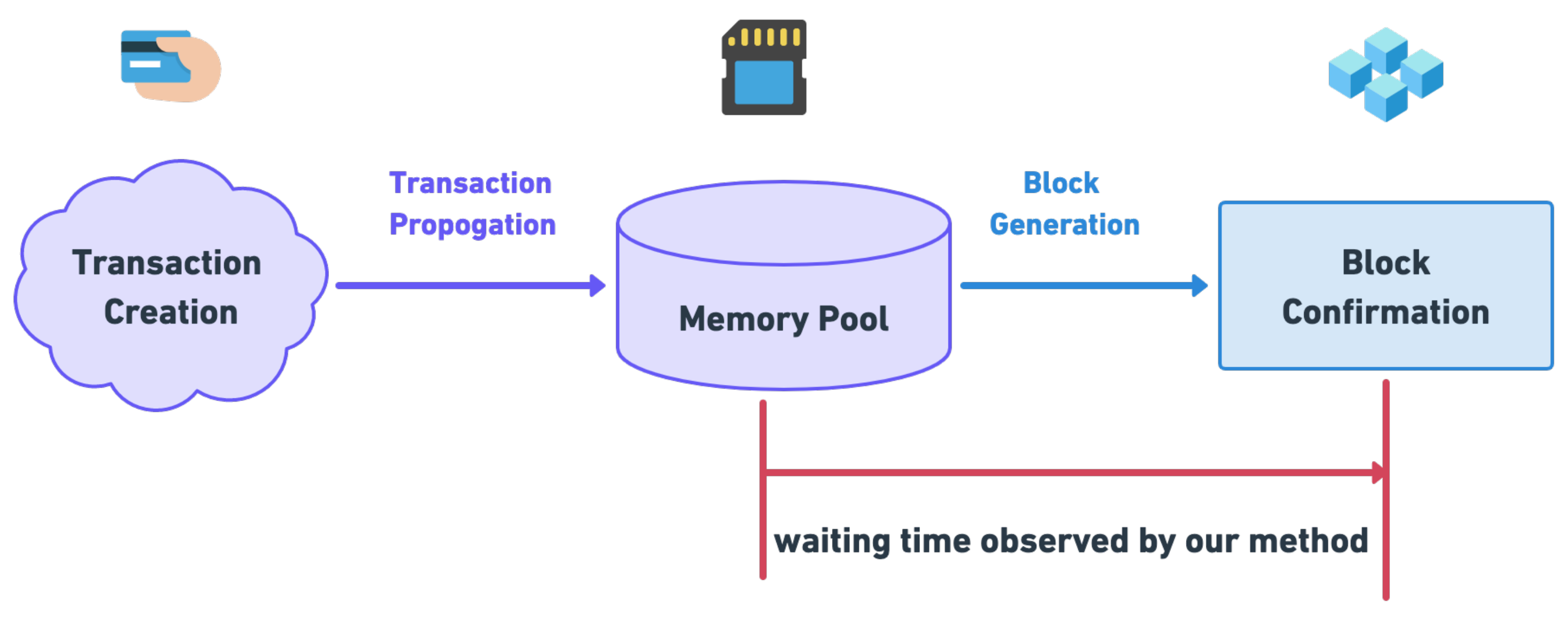}
	\caption{The life cycle of an Ethereum transaction through the public mempool and the definition of its waiting time.}
	\label{figWT}
\end{figure}

In blockchain applications, transaction waiting time---namely, the time a user must wait for her transactions to be mined---is essential, and reducing waiting time is generally desirable.
\Cref{figWT} illustrates the life cycle of Ethereum transactions through the public mempool.
In market design, transaction waiting time is an important metric of market congestion~\cite{hall2018pareto, mayer2003network, anderson2014subways}---a higher waiting time indicates a more congested market of lower liquidity and allocation efficiency~\cite{roth2008have}. More importantly, earlier theoretical research on TFM highlights the importance of no market congestion condition for other desired outcomes that also matters for blockchain security: \citet{roughgarden2020transaction,Roughgarden2021} prove that under no market congestion, or in the word of theoretical language---when the block size is infinite, the current TFM on Ethereum called EIP1559 approximate a "burning posted price" auction that achieves all three properties of user incentive compatibility (UIC), miner incentive compatibility (MIC), and c-side-contract-proof (c-SCP) at the same time. However, when the market is congested, or the block size is finite, we face a general trilemma where all three properties are unlikely to be satisfied simultaneously. \citet{chung_2022_foundations} first document a finite-block impossibility theorem under conventional settings. \citet{shi2022can} endeavor to circumvent the impossibilities by exploring two directions---one allows for a weaker version of incentive compatibilities, and the other employs cryptography to equip TFM with MPC-assisted models. However, \citet{shi2022can}'s results further echo the finite-block impossibility theorem. How to make the impossible possible? \citet{roughgarden_2020_transaction,Roughgarden2021,chung_2022_foundations,shi2022can} inspire that understanding waiting time and avoiding market congestion is crucial in achieving desired outcomes in TFM. 

However, in practical systems, transaction waiting time is a function of several intertwined factors, including system parameters (e.g., block interval and block size)~\cite{Yaish2022EC, alharby2023transaction}, network latency~\cite{tang_2022_strategic}, the level of congestion~\cite{lotem2022sliding} and so forth, which makes it challenging to understand the exact mechanisms behind observed waiting time. For instance, a recent empirical study~\cite{liu2022empirical} suggests the London fork~\cite{ethereumfoundation_2021_london}, a major upgrade on the Ethereum blockchain that implemented the Ethereum Improvement Proposal 1559 (EIP-1559) \cite{ethereumimprovementproposals_2021_ethereum} on August 5th, 2021, significantly reduced waiting time as an unintended result. However, the existing theoretical framework~\cite{roughgarden_2020_transaction,Roughgarden2021, chung_2022_foundations,shi2022can} cannot explain those phenomena. 

Do policy shocks on blockchain affect transaction waiting time as Gettier cases~\cite{blouw2018gettier} of a single incidence or repetitive phenomenon worth theorizing? Moreover, \citet{reijsbergen2021transaction} shows that the EIP-1559 adjusts the base fee slowly during periods of demand bursts of market congestion, which is likely to be due to shocks such as Non-Fungible Token (NFT)~\cite{wang2021non, nadini2021mapping,vasan2022quantifying} drops. Then, can we improve the forecast of demand surges considering community events such as NFT drops? If we could efficiently predict demand bursts and design TFM to coordinate the market better, we could prevent market congestion from happening ex-ante. Our research aims to answer the following three questions seeking the impacts of shocks on transaction waiting time: 

\begin{enumerate}
    \item \textbf{Ceteris Paribus:\footnote{Ceteris paribus is a Latin phrase generally meaning "all other things being equal." In economics, it acts as a shorthand indication of one economic variable's effect on another, provided all other variables remain the same~\cite{reutlinger2011ceteris}. Refer to~\url{https://dictionary.cambridge.org/dictionary/english/ceteris-paribus}.}} How does the Merge, the most recent significant upgrade of Ethereum, affect the latency in the EIP-1559 transaction fee mechanism? In the meantime, \textit{Ignorantia juris non excusat,\footnote{In law, ignorantia juris non excusat---latin maxim for ignorance of the law excuses not---is a legal principle holding that a person who is unaware of a law may not escape liability for violating that law merely by being unaware of its content~\cite{keedy1908ignorance}.}} what are the other unobservables and confounding factors that might affect the latency in EIP-1559 TFM?
    \item \textbf{Ad Infinitum:\footnote{Ad Infinitum is used to designate a property that repeats in all cases in mathematical proof and also used in philosophical contexts to mean "repeating in all cases"~\cite{yoshida2008game}. Refer to~\url{https://dictionary.cambridge.org/dictionary/english/ad-infinitum}.}} Can we abstract a mathematical model that explains the major impact of the Merge on TFM in particular and guides future TFM Design in general? 
    \item \textbf{Mutatis Mutandis:\footnote{Mutatis mutandis is a medieval Latin phrase meaning "with things changed that should be changed" or "once the necessary changes have been made"~\cite{stoyle2005mutatis}. Refer to \url{https://dictionary.cambridge.org/dictionary/english/mutatis-mutandis}.}} How do NFT drops interact with market congestion in the EIP-1559 transaction fee mechanism?
\end{enumerate}

\subsection{Challenges and Our Approach}
Our research is challenging in synthesizing the insights from computing and economic science applying to the TFM on the Ethereum blockchain in facets of research questions, methodologies, and application scenarios. 
\begin{enumerate}
    \item \textbf{research questions}: Although deeply connected, the waiting time issues are intertwined with different terms in computing and economic science. To recognize the interdisciplinary connection requires the understanding of both literature fundamentally. We review and connect related terms across disciplines and provide a glossary table in \Cref{appendix:glossary table} that lists several examples in this regard.
    \item \textbf{methodologies}: The methodology for drawing inferences on both facets of security and efficiency from blockchain and mempool data does not exist. We match the right tool to the jobs of \textit{ceteris Paribus} and \textit{mutatis mutandis}, respectively. First, to understand the causal inference of the Merge, we apply the Regression Discontinuity Design (RDD) method ~\cite{thistlethwaite1960regression}. RDD, widely applied to economics and other social sciences, is a quasi-experimental pretest-posttest design able to estimate the local average treatment effect (ATE) of natural experiments such as the Merge when a counterfactual generated by randomized controlled trials do not exist~\cite{imbens2008regression}. Second, to study the interaction between NFT drops and market congestion, we use Facebook Prophet~\cite{taylor2018forecasting}, an open-source algorithm for generating time-series models that decompose the forecast into components of trend, seasonal, and holiday effects. We also use the Python NetworkX \cite{networkx} to compare the networks for specific transactions that might affect the results. 
    \item \textbf{application scenarios}: Applying standard causal inference methods to estimate the effect of the Merge faces challenges because of the complexity of blockchain systems and the several confounding factors that affect waiting time. For example, the RDD method does not automatically reject causal effects by potential confounding variables that by coincidence have a treatment effect~\cite{imbens2022causality} on the waiting time at the time of the Merge. \citet{cunningham2021causal} points out that no general test exists for excluding the effects from confounding factors, for which researchers must exercise domain expertise to evaluate the causal inference. \citet{yang2022sok} shows that the transactions sanctioned by the U.S. Treasury's Office of Foreign Asset (OFAC)~\cite{arnold2022stolen} tend to have higher waiting times. We consider the potential confounding effect by examining all the sanctioned transactions during our data range. 
\end{enumerate}

\subsection{Our findings}
\subsubsection{Ceteris Paribus}
We find that the Merge reduces the high risk of long waiting time, network loads, and market congestion on Ethereum despite that the transaction arrival rate---the average number of new transactions submitted to the public mempool per second---increases from 12.1 to 13.0 transactions per second. 
.  

\begin{itemize}[leftmargin=*]
    \item {\em Waiting time.} 
    The Merge significantly decreases the upper quantile---e.g., the local ATE of the Merge on the 75\% quantile is a (38\%) reduction of 13.4 seconds from 35.0 seconds to 21.6 seconds. Moreover, the intrablock waiting time, measured by the Interquartile range (IQR), becomes significantly lower after the Merge---the local ATE of the Merge on the IQR is a (47.8\%) reduction of 26.1 seconds from 54.5 seconds to 28.4 seconds. The p-values of the results---error rates of rejecting the null hypothesis that the Merge has no ATE---are all below 0.01.{\bf Our results suggest that with everything else equal, user transactions are included in a block much quicker on average after the Merge. This could improve user experience and reduces the risk of long waiting time when trading on Ethereum.}
    
    \item {\em Network load.} Following~\citet{liu2022empirical}, we define the network load as the amount of computational, networking, and storage work that a node must perform to participate in the blockchain protocol, measured by, for instance, in our paper, the gas used per second in Ethereum. A significant increase in the network load could negatively affect the security of blockchains, as it prevents computationally weak nodes from participating---leading to centralization, and increases block processing time--leading to higher fork rates~\cite{garay2015bitcoin, eip1559vitalik}. {\bf Our results indicate that the Merge significantly reduces the network load of the Ethereum system, which positively influences blockchain security.} For instance, the ATE of the Merge on the 1-block, 5-block, and 7200-block moving average network loads are a reduction of 28.55\%, 28.50\%, and 22.31\%, respectively. The p-values of the results---error rates of rejecting the null hypothesis that the Merge has no ATE---are all below 0.01. 
    
    \item {\em Market congestion.} We define market congestion as when a block consumes more than 95\% of the block gas limit. Using a Logit Regression~\cite{christodoulou2019systematic} with RDD, we find that the Merge significantly reduces the relative risks---i.e., the odds ratio of the probability of being congested and not being congested---of market congestion by 52.72\%. Moreover, the Merge also significantly decreases the relative risk that the market is congested for 5 continued blocks ($\sim$ 60 seconds) by 41.08\%. The p-values of the results---error rates of rejecting the null hypothesis that the Merge has no ATE---are all below 0.01. {\bf Our result implies that the Merge considerably improves market efficiency by reducing the possibility of being (persistently) congested.} \textit{Ignorantia juris non excusat}: we reason that other factors such as the unobserved delay and systematic transaction preference such as that caused by OFAC sanctions are not likely to affect our results.    
\end{itemize}


\subsubsection{Ad Infinitum}
After examining three major protocol changes of the Merge, we find that our results are most likely to be the direct consequence of the block interval change. We abstract a mathematical model to elaborate that the base fee adjustment in EIP1559 is crucial for the Merge, precisely the block interval reduction to positively affect waiting time, network loads, and market congestion simultaneously. {\bf A reduction in block interval drives the base fee in EIP1559 to adjust much faster during any demand surge so as to provide more prompt signals for the fee market on Ethereum.} Thus, our result guides designing the block interval for future TFM. 

\subsubsection{Mutatis Mutandis}

We find that considering the shock of NFT drops significantly improves the forecast of market congestion, especially for persistent market congestions. We define the congestion ratio as the intraday percentage of congested blocks. And we define the continued congestion ratio as the intraday percentage of blocks being continued congested for 5 blocks. {\bf The time-series model successfully identifies the NFT drops as holiday effects besides the trend and seasonal effects. Moreover, the NFT drops successfully match the peaks in (continued) congestion ratios}.    


 \subsection*{}
 The rest of the paper is structured as follows. \Cref{sec: literature} discusses related literature. \Cref{sec: data and method} introduces the data and methods. \Cref{sec: merge} presents the results for the empirical analysis of the merge and examines the confounding factors. \Cref{sec: model} models block interval as a new design feature for EIP-1559 TFM. \Cref{sec: NFT} evidences NFT as a typical case of demand surge. In \Cref{sec:future TFM}, we envision the future of TFM design. Our open-source data and code are available on GitHub: \url{https://github.com/cs-econ-blockchain/waiting-time-eip1559}.
 
 

\section{Related Literature}
\label{sec: literature}

Our research contributes to the literature on the interplay of computing and economic science around the applications of blockchain technology. 
\subsection{The contribution to literature in economics}
\begin{itemize}[leftmargin=*]
    \item {\em Market and Mechanism Design on Blockchain.} Mechanism Design Theory~\cite{mohanta2018overview}, initiated by 2007 Nobel prize laureates in economics---Leonid Hurwicz, Eric S. Maskin, and Roger B. Myerson designs---incentives towards desired objectives. The Nobel Prize laureates in 2012---Alvin E. Roth and Lloyd S. Shapley---apply mechanism theory to specific economic problems such as school choice and employment and lay the foundation for market design~\cite{masso2015theory}. However, existing mechanisms and market design theory require a trusted third party in execution, which is difficult to be guaranteed. In contrast, Ethereum Blockchain enables the TFM to be executed automatically in a distributed system. Thus, to guarantee the truthful execution of the TFM on Ethereum does not require a trusted third party but becomes a computer security problem~\cite{gollmann2010computer}. The pioneering literature that studies the mechanism and market design of TFM on blockchains, including~\cite{roughgarden_2020_transaction}, \cite{Roughgarden2021}, \cite{chung_2022_foundations}, \cite{liu2022empirical}. Our research on waiting time in TFM advances the literature by further seeking the desired outcomes of both economic and security importance. 
    \item {\em Causal Inference for Natural Experiments on Blockchain.} 2021 Nobel prize laureates in economics---David Card, Joshua D. Angrist, and Guido W. Imbens---apply causal inference to natural experiments to study the policy changes that affect human welfare~\cite{banerjee2020field}. However, before blockchain technology existed, meaningful data from natural experiments were rarely available for research. Blockchain technology makes the valuable data before and after the natural experiments on Ethereum automatically recorded and tamper-proof. The emerging literature applied the causal inference method to study the natural experiment of EIP-1559, including \citet{liu2022empirical}, \citet{zhang2022sok}, and \citet{cong2022inclusion}. Integrating blockchain data with ephemeral mempool data, our research expands the literature to study the policy effect on computer security and economic efficiency before and after the Merge, the most recent major updates of Ethereum. 
    \item {\em Time-series Analysis Forecast on Blockchain.} 2013 Nobel prize laureates in economics---Eugene F. Fama, Robert J. Shiller, and Lars Peter Hansen---apply time-series analysis to asset pricing so as to successfully forecast the price changes of stocks and bonds~\cite{shiller2014speculative}. An emerging literature extends the time-series analysis to the crypto-assets valuation on blockchain including~\citet{liu2022cryptocurrency}, \citet{liu2021risks}, and \citet{cong2021value}. We expand the application of time-series analysis to the forecast of market congestion in the TFM on Ethereum. 
\end{itemize}
\vspace{3mm}

\subsection{The contribution to~literature~in~ computer science}

\begin{itemize}[leftmargin=*]
    \item {\em EIP-1559 TFM.} Our research contributes to the emerging literature on blockchain TFM, specifically the EIP-1559 TFM on Ethereum~\cite{roughgarden_2020_transaction,Roughgarden2021, chung_2022_foundations,zhao2022bayesian}. In this regard, ours is most related to \citet{liu2022empirical}, which empirically studied how EIP-1559 affects the waiting time and consensus security of Ethereum in terms of fork rates, network loads, and the composition of miner revenue. This work extends the methodology of~\cite{liu2022empirical} to study the implications of the Merge.
    \item {\em Empirical Analysis of the Merge.} 
    Our paper presents empirical measurements of the impacts of the Merge.
    Ours is mostly close to \citet{kapengut2023event} that compares various metrics, such as energy consumption, miner rewards, token supplies, transaction fees, and so forth, before and after the Merge. However, \citet{kapengut2023event} does not aim to understand the causal inference of the Merge as ours. Neither did \citet{kapengut2023event} study transaction waiting time nor network loads as ours, although touching on block creation time. 
    \item {\em Empirical security of Post-merge Ethereum.} 
    The consensus security---or attacks---of the post-Merge Ethereum protocol, has been extensively studied---see, e.g., \citet{d2022no,pavloff2022ethereum} and the references therein. We study the impact of the Merge on network loads whose implications on security are independent of the specific consensus mechanisms.
\end{itemize}
\vspace{3mm}



\section{Data and Method}
\label{sec: data and method}
\subsection{Data}

\subsubsection{Data Source}
Our data source includes the Ethereum blockchain and mempool data as well as the OFAC-sanctioned addresses for Tornado Cash censorship.~\footnote{https://home.treasury.gov/news/press-releases/jy0916} \cref{tab12} in~\Cref{sec: data dictionary} shows the data dictionary for queried variables. 
\begin{itemize}[leftmargin=*]
    \item {\em Ethereum blockchain data}: We use ethereum-etl~\footnote{https://github.com/blockchain-etl/ethereum-etl}---python scripts for extract, transform and load (ETL) jobs for Ethereum---to query the blockchain data. We used the endpoints provided by QuickNode~\footnote{https://www.quicknode.com/} with a paid subscription. We queried both the block and the transaction data. 
    \item {\em Ethereum mempool data}: We use the mempool data provided by the Mempool Guru service\footnote{https://mempool.guru} that uses multiple full nodes to monitor the Ethereum mempool and record the time stamp when transactions first appear in the mempool. Mempool Guru also records the time when a block is first observed by our full nodes. The difference between the two timestamps is the waiting time of a given transaction, with delay as the variable name in the database. The set difference between transactions in the blockchain and those in the mempool is {\em private transactions} that are included in the blockchain without first entering the public mempool. 
    \item {\em OFAC-sanctioned address}: We obtained the list of OFAC-sanctioned addresses for Tornado Cash censorship from the OFAC program website.~\footnote{https://home.treasury.gov/policy-issues/office-of-foreign-assets-control-sanctions-programs-and-information} Because our data range ends on Sep. 25th, we choose the list before the Nov. 8th update. 
\end{itemize}
\subsubsection{Data Scope}
Our data consists of two ranges for studying the Merge and the NFT drops, respectively.  
\begin{itemize}[leftmargin=*]
    \item {\em The Merge}: \cref{tab13} represents the data range before and after the Merge for a total of 140,000 blocks. 
    \item {\em NFT drops}: \cref{tab14} shows the data range covering three NFT drops including Fatales on Aug.31, 2021, Pointilla on Sep.9, 2021, and GalaxyEggs on Sep.14, 2021, for a total of 181,075 blocks. 
\end{itemize}
\begin{table}[!htbp]
  \caption{The data range before and after the Merge: total block numbers$=$140,000, total transactions$=$23,871,253 }
    \centering
    \begin{tabular}{|l|l|l|}
    \hline
        type & from & to \\ \hline
        date\_time & 2022-09-03 20:10:39 & 2022-09-25 02:27:11 \\ \hline
        unix\_timestamp & 1662235839.0 & 1664072831.0 \\ \hline
        block number & 15467393 & 15607393 \\ \hline
    \end{tabular}
    \label{tab13}
\end{table}
\begin{table}[!htbp]
  \caption{The data range covers three NFT drops: total block numbers$=$181075}
    \centering
    \begin{tabular}{|l|l|l|}
    \hline
        type & from & to \\ \hline
        date\_time & 2021-08-24 00:00:00 & 2021-09-21 00:00:00 \\ \hline
        unix\_timestamp & 1629763200.0 & 1632182400.0 \\ \hline
        block number & 13084679 & 13265754 \\ \hline
    \end{tabular}
    \label{tab14}
\end{table}

\subsection{Method}
\subsubsection{Regression discontinuity design (RDD)}
We apply RDD \cite{thistlethwaite1960regression} to estimate the local ATE of the Merge on waiting time, network load, and market congestion. We specify the RDD by~\cref{eqn:reg}: 

\begin{equation}\label{eqn:reg}
Y = \alpha_0 + \alpha_1 \mathbbm{1}(\text{merge}) + \mathbf{\alpha_2 X} +\mathbf{\alpha_3} \mathbbm{1}(\text{merge})\mathbf{X}+ \epsilon.
\end{equation}

Here,  $\alpha_1$ is the coefficient for the indicator variable for the occurrence of the Merge (affecting block number $\geq$ 15537393), which characterizes the average treatment effect of the Merge. $\alpha_2$ is the coefficient for the control variable $\mathbf{X}$. We control for the block number in our sample to account for a possible time trend before the Merge, defined by %
\[
\text{blockn} =
\begin{cases}
 \text{number} - 15537393 & \text{pre-merge period}  \\ 
 \text{number} - 15537393 & \text{post-merge period}
\end{cases}
.
\]

$\alpha_3$ is the coefficient for the intersection term $\mathbbm{1}(\text{merge})\mathbf{X}$, which captures the average treatment effect of the Merge interacting with the time trend. 

\subsubsection{Time series forecast}
We apply Facebook Prophet~\cite{taylor2018forecasting}, an open-source algorithm~\footnote{https://github.com/facebook/prophet} to generate decomposable time-series models~\cite{harvey1990estimation}. We specify the model by~\cref{eqn:decompose}.

\begin{equation}
\label{eqn:decompose}
y(t) = g(t) + s(t) + h(t) + \epsilon(t).
\end{equation}
Here $g(t)$ is the trend function that models non-periodic changes in the value of the time series, $s(t)$ represents weekly periodic changes, and $h(t)$ represents the effect of holidays that occurs on potentially irregular schedules for the days of NFT drops. The error term $\epsilon(t)$ represents any idiosyncratic changes that are not accommodated by the model assumed to be normally distributed. 
\subsubsection{Network analysis}
We use Python NetworkX \cite{networkx}, an open-source package for network analysis, to draw the undirect graph~\cite{yang2012feature} of OFAC-sanctioned transactions before and after the Merge. 

\section{An Empirical Analysis of the Merge}
\label{sec: merge}

In this section, we present our findings. Then we discuss the implications in the next section.
\subsection{Results on Waiting Time, Network Load, and Market Congestion}
\label{sec: merge_results}
\textbf{Per Ceteris Paribus}: how does the Merge, the most recent significant upgrade of Ethereum, affect the latency in the EIP-1559 transaction fee mechanism?
We find that the Merge remarkably reduces the risk of the high opportunity cost of time, network loads, and market congestion on Ethereum despite that the transaction arrival rate---the average number of new transactions submitted to the public mempool per second---increases from 12.079 to 12.997.

\subsubsection{Waiting time}
\begin{figure}
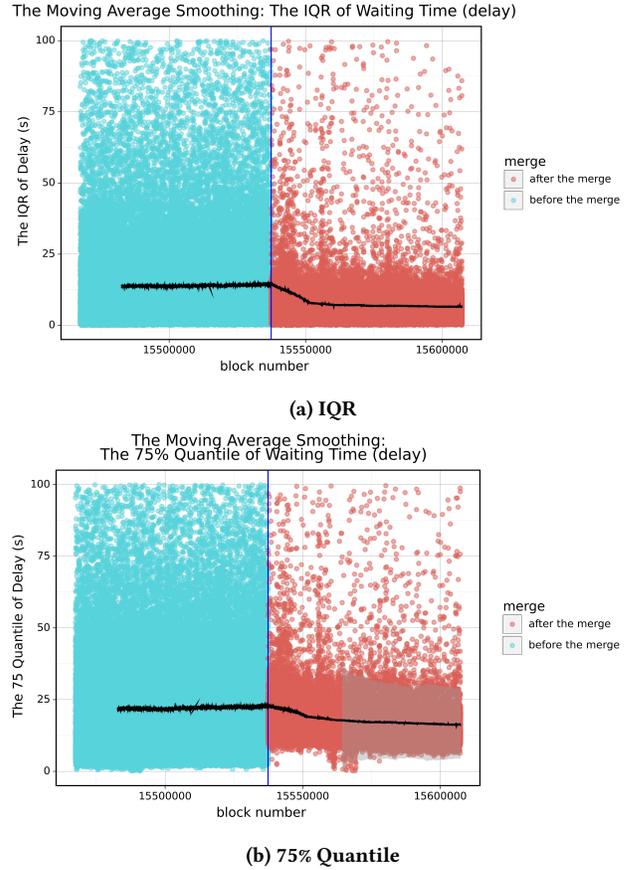

\begin{subfigure}{\columnwidth}
  \centering
  \includegraphics[width=\columnwidth]{figs/fig13.pdf}
  \caption{IQR}
  \label{fig13}
\end{subfigure}
\begin{subfigure}{\columnwidth}
  \centering
  \includegraphics[width=\columnwidth]{figs/fig14.pdf}
  \caption{75\% Quantile}
  \label{fig14}
\end{subfigure}
\caption{Moving Average Smoothing for Waiting Time (delay) by Block Before and After the Merge}
\label{fig:Moving Average Smoothing for Waiting Time (delay) by Block}
\end{figure}

We define waiting time as the difference between the time when we first observe the transaction in the mempool and when the transaction is mined. \Cref{fig:Moving Average Smoothing for Waiting Time (delay) by Block} shows the moving average smoothing for the intrablock  IQR and the 75\% quantile of the waiting time before and after the Merge. In~\Cref{sec: regression tables}, \Cref{tab4} and \Cref{tab5} are the RDD results for the intrablock 75\% quantile and IQR of the waiting time around the Merge. The Merge significantly decreases the upper quantile---the local ATE of the Merge on the 75\% quantile is a reduction of 13.4 seconds from 35.0 seconds to 21.6 seconds as shown in \Cref{tab4} Column (3). The p-values of the results---error rates of rejecting the null hypothesis that the Merge has no ATE---are all below 0.01. Moreover, the intrablock waiting time, measured by the Interquartile range (IQR), becomes significantly lower after the Merge---the local ATE of the Merge on the IQR is a reduction of 26.1 seconds from 54.5 seconds to 28.4 seconds as shown in \Cref{tab5} Column (3). The p-values of the results---error rates of rejecting the null hypothesis that the Merge has no ATE---are all below 0.01. Our results imply that the Merge improves user experience remarkably by reducing the risk of high opportunity cost to transact on the Ethereum blockchain.

\subsubsection{Network load}
Prior to the Merge, Ethereum block production is a stochastic process and the block interval roughly follows an exponential distribution. When blocks are produced in quick succession, nodes will need to spend more computation and bandwidth to process blocks, experiencing load bursts.
\citet{liu2022empirical} quantified the distribution of network loads, defined as the average gas per second in a given time period. \citet{liu2022empirical} show that while the expectation is about 1.15 million gas/second, the highest observed network loads can be much higher (e.g., 6 million gas/second). 
After the Merge, however, blocks are produced at a constant rate (every 12 seconds). A natural question is how would that affect network loads. 

\Cref{fig:Moving Average Smoothing for Waiting Time (delay) by Block} shows the histograms of 5-block and 7200-block moving average gas used per second before and after the Merge. \Cref{tab8}, \Cref{tab9}, and \Cref{tab20} in \Cref{sec:statistics} print the statistics of gas used per second, the 5-block moving average of gas used per second, and the 7200-block moving average of gas used per second before and after the Merge. \Cref{tab10}, \Cref{tab11}, and \Cref{tab22} in \Cref{sec: regression tables} are the RDD results respectively. ATE of the Merge on the 1-block, 5-block(a period usually a bit above 60 seconds), and 7200-block (a period usually a bit above 1 day) moving average network loads are a reduction of 28.55\%,~\footnote{from 1777314.312 to 1269919.127 by a reduction of 507395.185 units per second} 28.50\%,~\footnote{from 1776803.682 to 1270499.904 by a reduction of 506303.778 units per second} and 22.31\%~\footnote{from 1771125.786 to 1380930.813 by a reduction of 390194.973 units per second} respectively as calculated from \Cref{tab10}, \Cref{tab11}, and \Cref{tab22} Column (3). The p-values of the results---error rates of rejecting the null hypothesis that the Merge has no ATE---are all below 0.01. Thus, we conclude that the Merge significantly reduces the intrablock gas used per second controlling other confounding factors. 

\begin{figure}
\begin{subfigure}{\columnwidth}
\includegraphics[width = \linewidth]{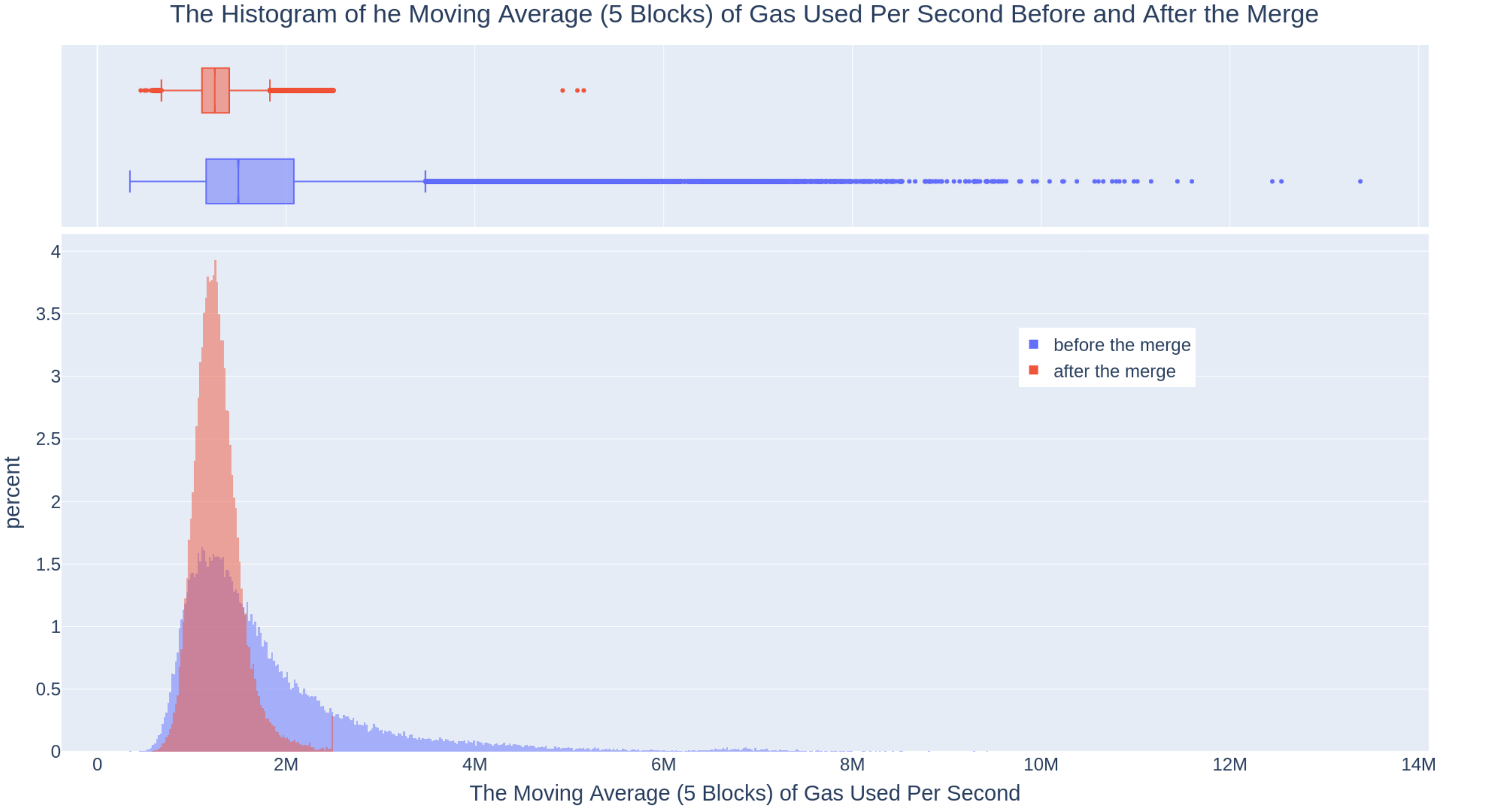}
\caption{5-block Moving Average}
\label{figLoad5}
\end{subfigure}
\begin{subfigure}{\columnwidth}
	\includegraphics[width = \linewidth]{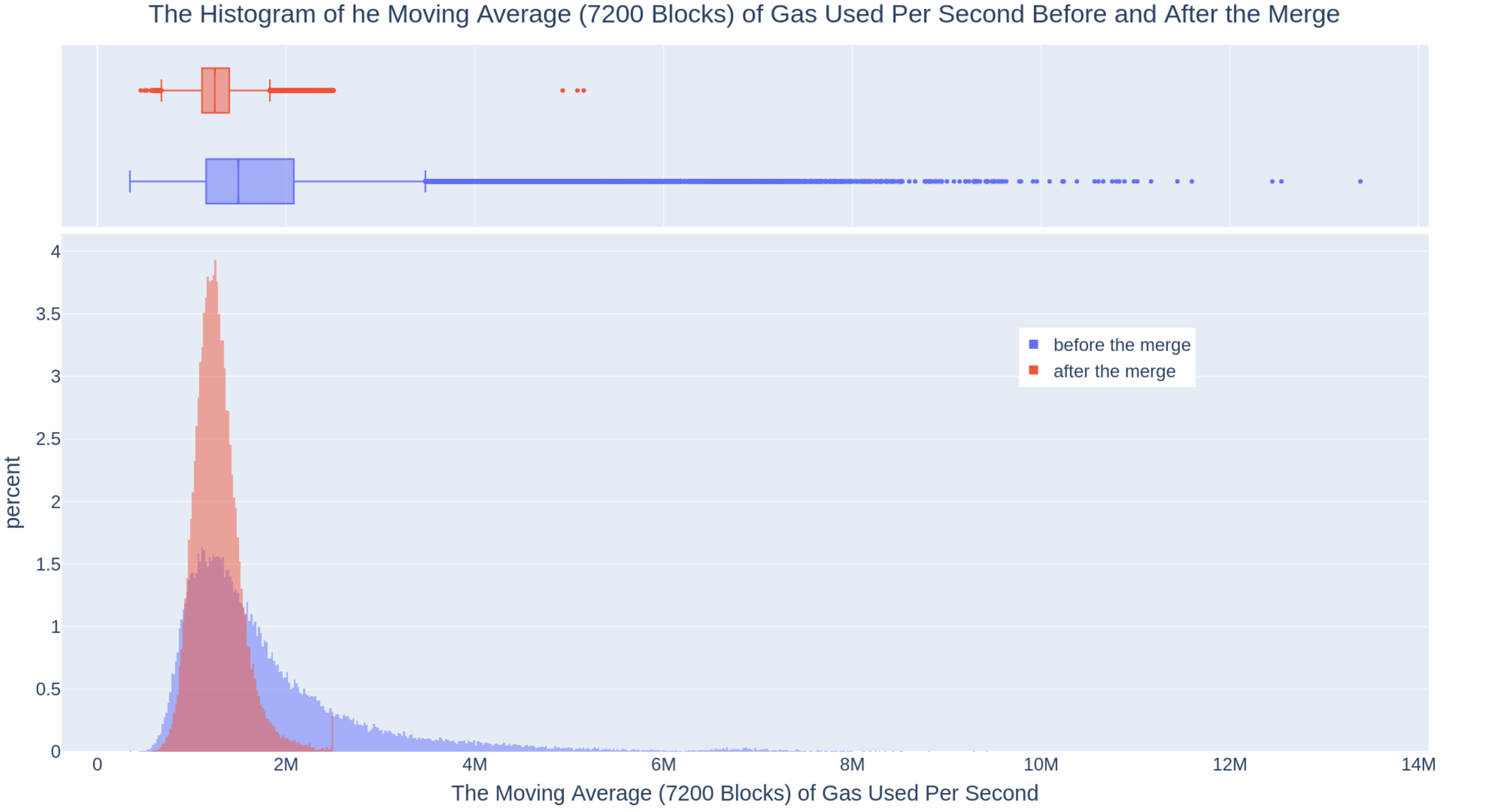}
	\caption{7200-block Moving Average}
	\label{figLoad7200}
\end{subfigure}
\caption{The Histogram of the Moving Average of Gas Used Per Second Before and After the Merge}
\label{fig:The Histogram of the Moving Average of Gas Used Per Second Before and After the Merge}
\end{figure}

\subsubsection{Market congestion.}
\begin{figure}
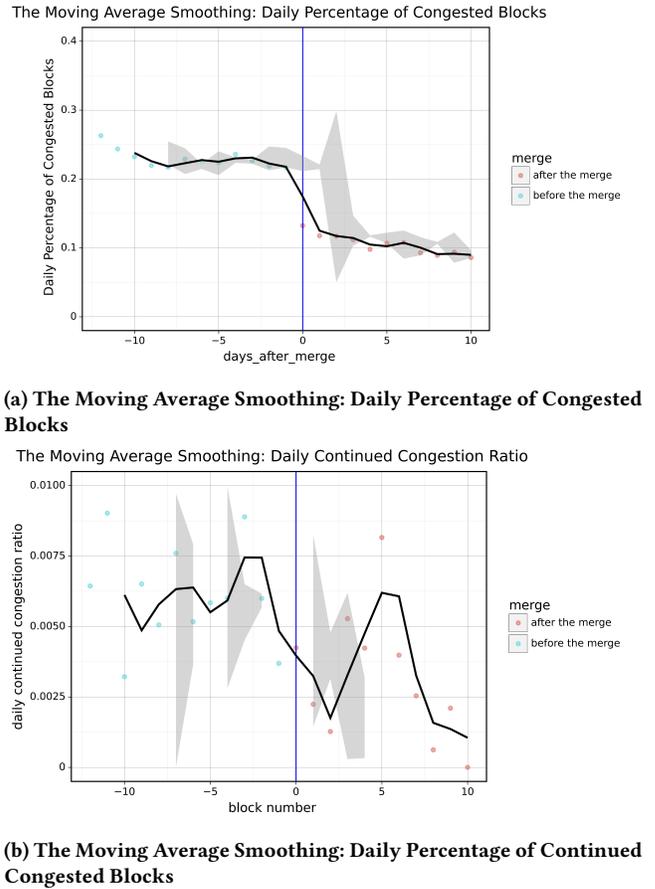

\begin{subfigure}{\columnwidth}
	\includegraphics[width = \linewidth]{figs/fig17.pdf}
	\caption{The Moving Average Smoothing: Daily Percentage of Congested Blocks}
	\label{fig17}
 \end{subfigure}
\begin{subfigure}{\columnwidth}
	\includegraphics[width = \linewidth]{figs/fig18.pdf}
	\caption{The Moving Average Smoothing: Daily Percentage of Continued Congested Blocks}
	\label{fig18}
\end{subfigure}
\caption{Daily Market Congestion Before and After the Merge}
\end{figure}

We define market congestion as when a block consumes more than 95\% of the block gas limit. \Cref{fig17} and \Cref{fig18} are the moving average of the daily percentage of congested and continued congested blocks. In~\Cref{sec: regression tables}, \Cref{tab1} and \Cref{tab2} are the RDD results using a Logit Regression~\cite{christodoulou2019systematic}. We find that the Merge significantly reduces the relative risks---i.e., the odds ratio of the probability of being congested and not being congested---of market congestion by 52.72\% as calculated by \Cref{tab1} Column (3).~\footnote{$1-{\rm e}^{-0.749}=52.72\%$} Moreover, the Merge also significantly decreases the relative risk that the market is congested for 5 continued blocks ($\sim$ 60 seconds) by 41.08\% as calculated by \Cref{tab2} Column (3).~\footnote{$1-{\rm e}^{-0.529}=41.08\%$} The p-values of the results---error rates of rejecting the null hypothesis that the Merge has no ATE---are all below 0.01.

\begin{figure}
	\includegraphics[width = \linewidth]{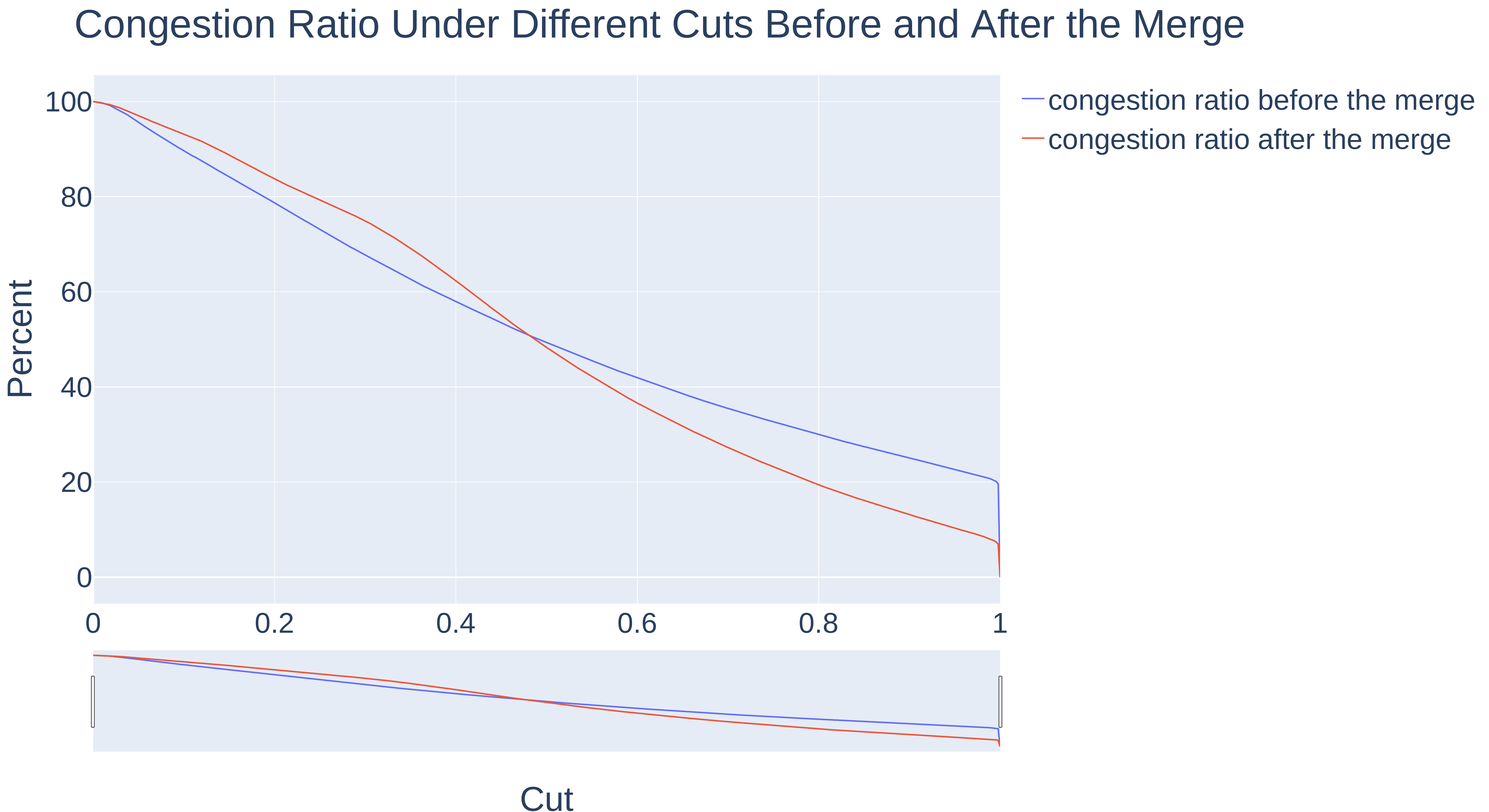}
	\caption{Congestion Ratio Under Different Cuts Before and After the Merge
 }
	\label{fig10}
\end{figure}

Thus, we conclude that We find the Merge significantly reduces the possibility of market congestion. Moreover, the Merge also significantly decreases the possibility that the market is congested for 5 continued blocks ($\sim$ 60 seconds). Our result implies that the Merge considerably improves market efficiency by reducing the possibility of being (persistently) congested. To further check the robustness of our result, we print the congestion ratio under different cuts, including the 95\% chosen in our analysis, before and after the Merge in~\Cref{fig10}. We can see that the congestion ratio, defined using any cuts larger than 50\%, decreases after the Merge.  
\subsection{Examine Confounding Factors}
\label{sec: confounding factors}
\textbf{Ignorantia juris non excusat}: what are the other unobservables and confounding factors that might affect the latency in EIP-1559 TFM?
\subsubsection{Results}
\label{sec: confounding_results}

First, not all transaction waiting times can be measured. The primary reason is the existence of various MEV auction platforms~\cite{yang2022sok} which effectively creates a private channel for transactions to be directly sent to miners (or proposers) without going through the public mempool. Their waiting time is thus private knowledge to MEV auction platforms.
\Cref{tab15} shows that the number of unobserved delays increases while that of the observed delay decreases after the Merge. We have good reason to hypothesize that delay of transactions via private channels, paid to be prioritized, is lower. In this case, we expect our results that the Merge decreases the median and 75\% quantiles of the waiting time to be stronger if considering the unobserved delay. However, currently, there exists no way to measure the unobserved delay that can verify or falsify our hypothesis. The transaction waiting time via private channels may be higher because relays add extra processing time before a transaction reaches a proposer. We leave the verification for future research. 
\begin{table}
\caption{Observed and Unobserved Delay Before and After the Merge}
\begin{tabular}{lrr}
\toprule
delay &         observed &       unobserved \\
merge &           &         \\
\midrule
before     &  11947760 &  326561 \\
after     &  11058935 &  537997 \\
\bottomrule
\end{tabular}
    \label{tab15}
\end{table}

\begin{table}
\caption{Sanctioned and Unsanctioned Transactions Before and After the Merge}
\begin{tabular}{lrr}
\toprule
merge &        before &        after \\
sanctioned &          &          \\
\midrule
no         &  9712821 &  9120233 \\
yes          &     1007 &      466 \\
\bottomrule
\end{tabular}
    \label{tab16}
\end{table}
\begin{figure}[!htbp]
\begin{subfigure}{\columnwidth}
	\includegraphics[width = \linewidth]{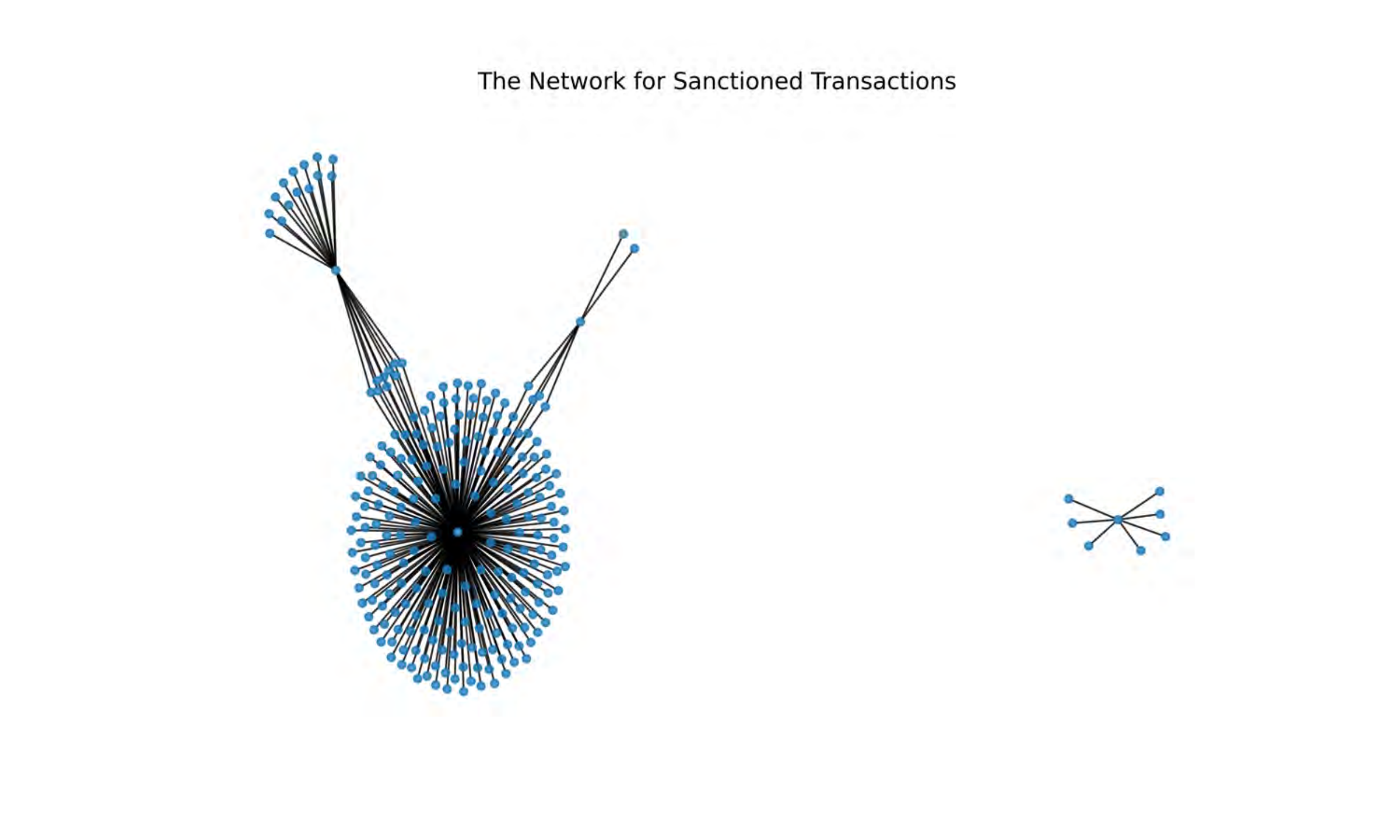}
	\caption{The Network for OFAC Sanctioned Transactions Before and After the Merge}
	\label{fig33}
\end{subfigure}
\begin{subfigure}{\columnwidth}
	\includegraphics[width = \linewidth]{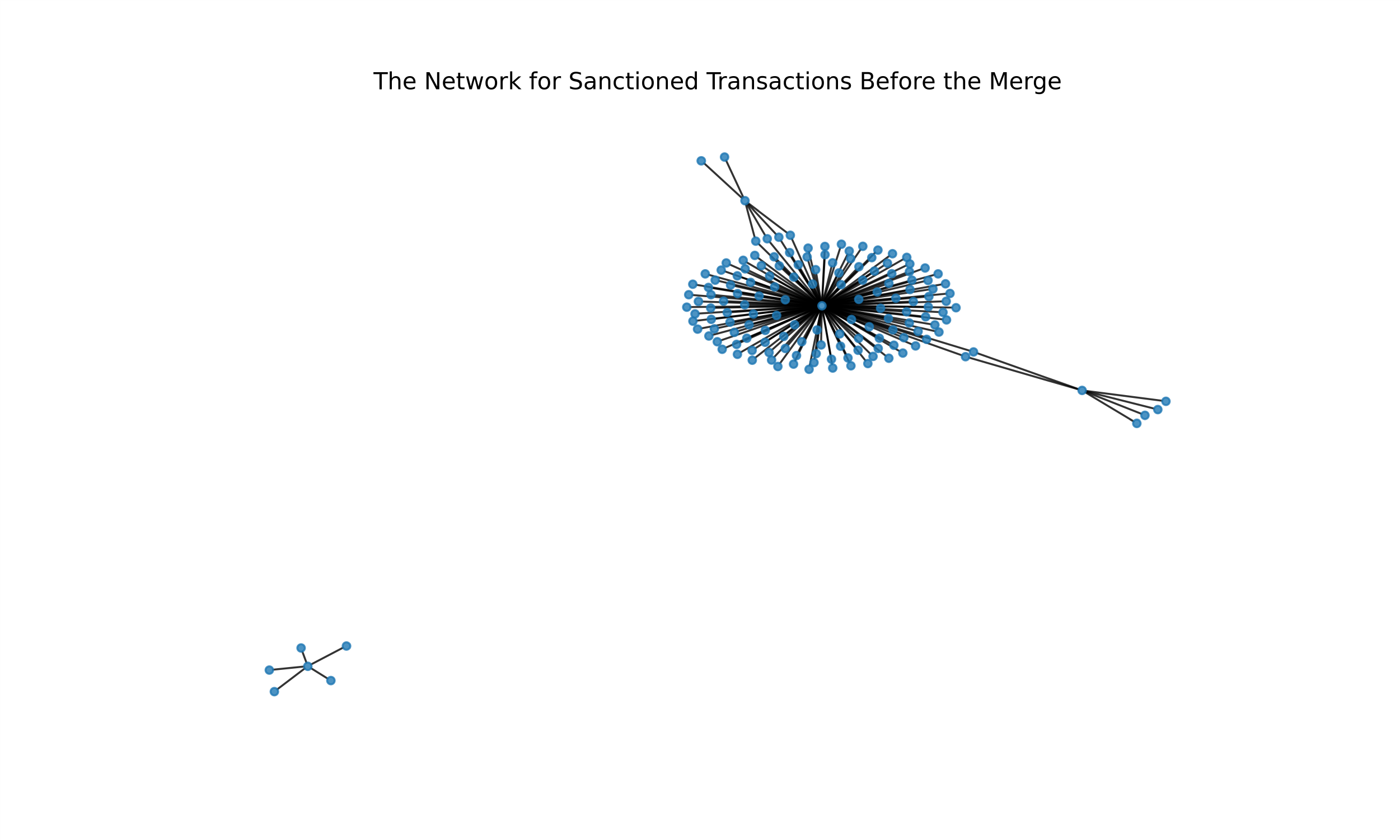}
	\caption{The Network for OFAC Sanctioned Transactions Before the Merge}
	\label{fig34}
\end{subfigure}
\begin{subfigure}{\columnwidth}
	\includegraphics[width = \linewidth]{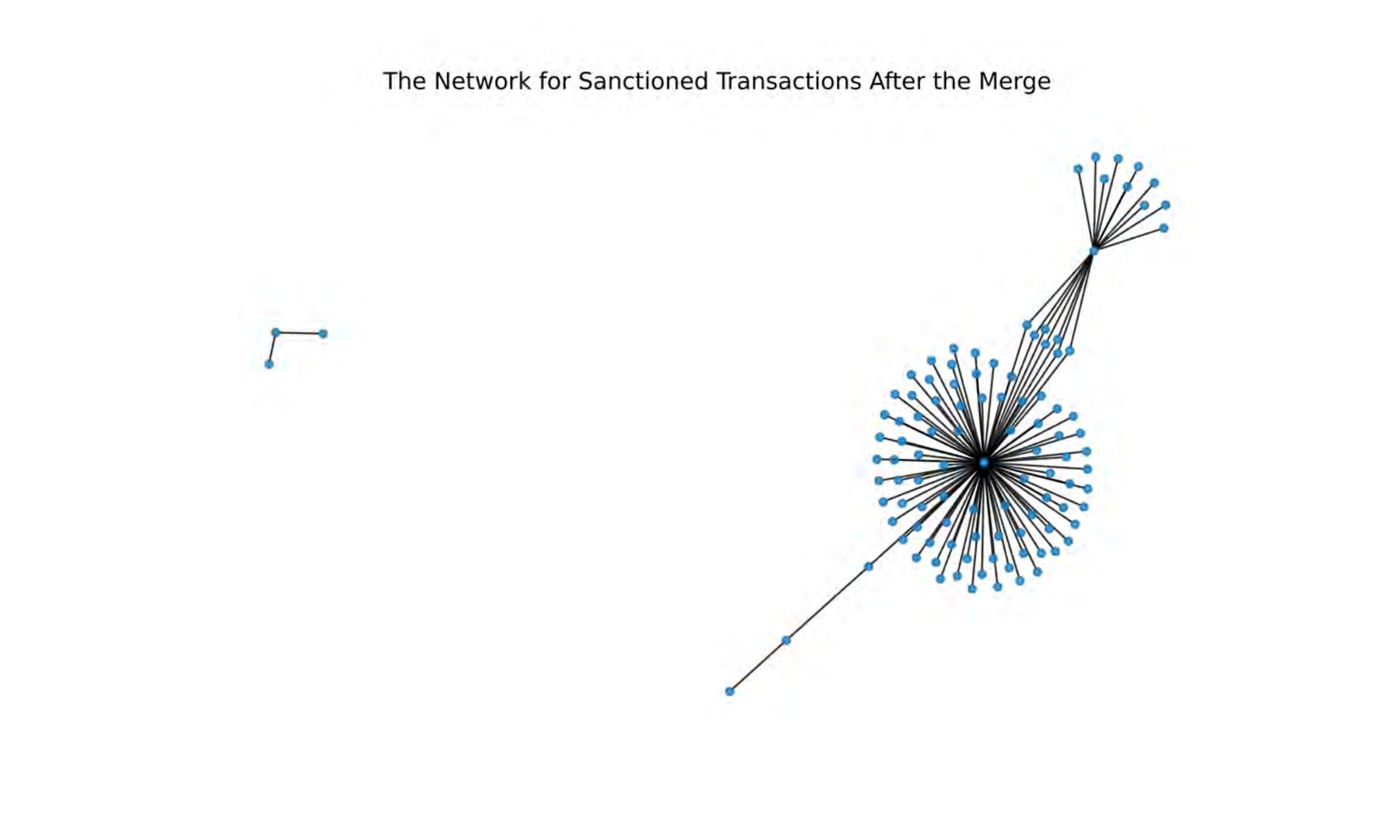}
	\caption{The Network for OFAC Sanctioned Transactions After the Merge}
	\label{fig35}
\end{subfigure}
\label{fig3345}
\caption{OFAC Sanctioned Transaction Networks}
\end{figure}

Another factor that may contribute to the waiting time is the existence of certain systematic preferences of transactions. For instance, the recent sanctions of Tornado Cash created such a preference where OFAC-compliant entities (e.g., relays, builders) will refuse to process offending transactions~\cite{yang2022sok}. As a result, the waiting time for those transactions will increase.
However, we first note from~\Cref{tab16} that the number of sanctioned transactions is much smaller compared to unsanctioned ones. Thus, their impact on the mean and quantiles of waiting time of {\em all} transactions should be minimal. \Cref{tab17} in the \Cref{sec: sanctioned transactions} shows the median and the 75\% quantile of delays decrease after the Merge for both the sanctioned and non-sanctioned transactions.
\Cref{fig33}, \Cref{fig34}, and \Cref{fig35} further document the network structure for OFAC-sanctioned transactions before and after the merge in undirected graphs.~\footnote{\url{https://networkx.org/documentation/stable/reference/classes/generated/networkx.Graph.to_undirected.html}} The edges are the sanctioned transactions and the vertices of the edge is either the sender or the receiver of the sanctioned transaction. The graphs suggest that the network structure for OFAC-sanctioned transactions does not change significantly before and after the Merge in undirected graphs. Thus we conclude that considering the confounding factor of OFAC-sanctioned transactions is not likely to change the causal inference of the Merge on waiting time statistics as it has neither structure changes nor affects a certain percentage of transactions around the merge. 

\begin{figure}[!htbp]
\begin{subfigure}{\columnwidth}
	\includegraphics[width = \linewidth]{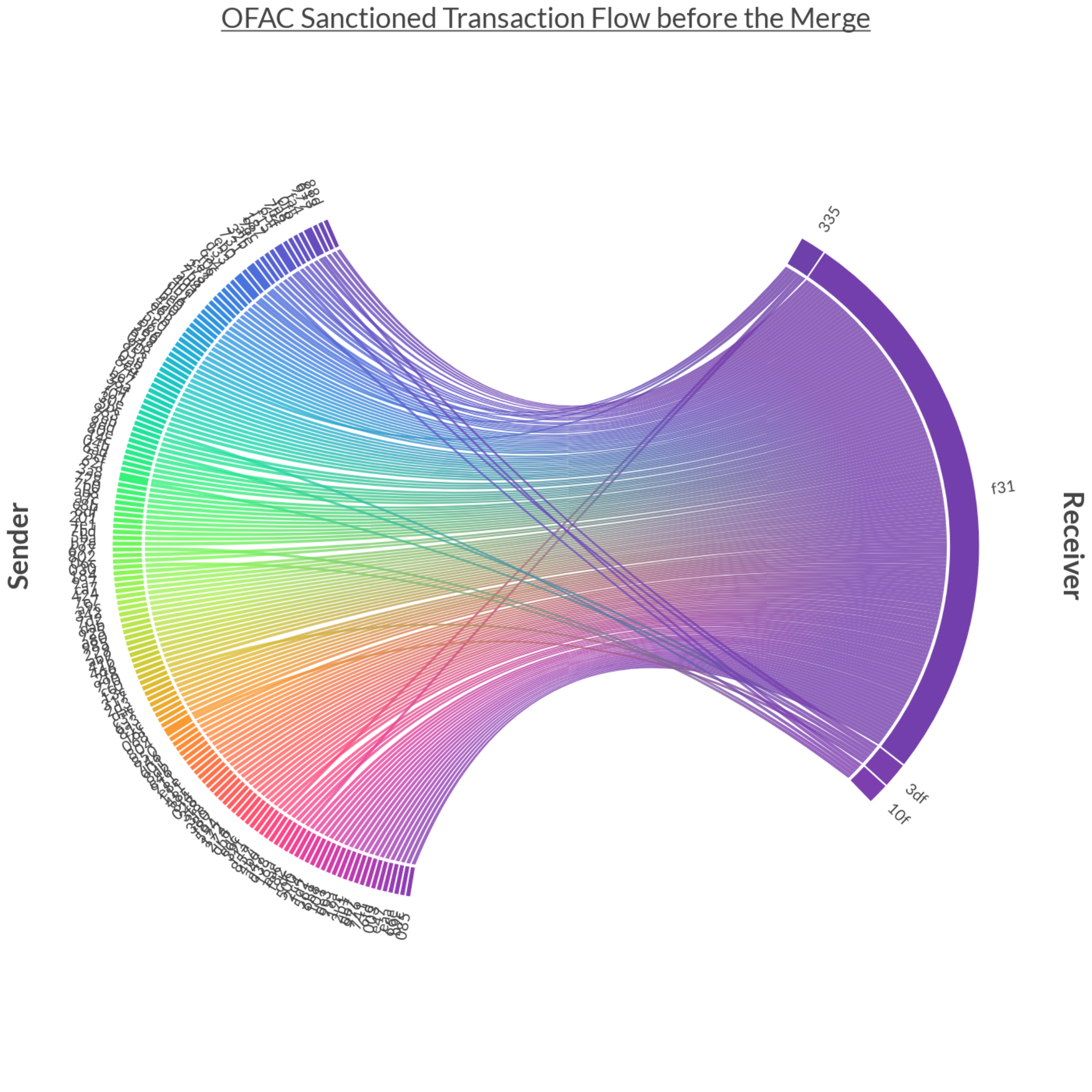}
	\caption{The OFAC Transaction Flow before the Merge}
	\label{fig: chord_before}
\end{subfigure}
\begin{subfigure}{\columnwidth}
	\includegraphics[width = \linewidth]{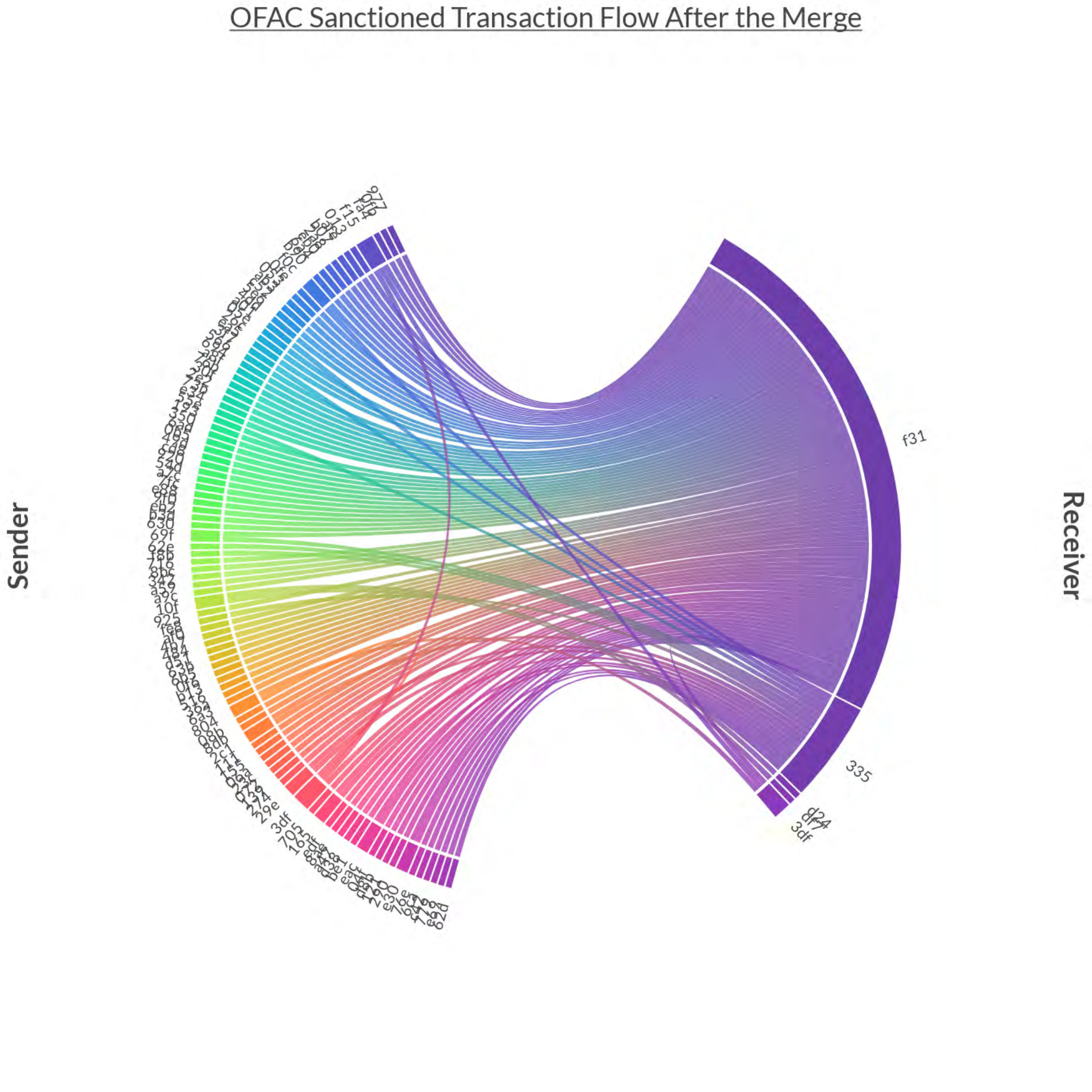}
	\caption{The OFAC Sanctioned Transaction Flow after the Merge}
	\label{fig: chord_after}
\end{subfigure}

\caption{The OFAC Sanctioned Transaction Flow before and after the Merge}
\label{fig:chord}
\end{figure}

We also find an interesting side result. First, \Cref{fig:chord} compares the OFAC Sanctioned Transaction Flow before (in \Cref{fig: chord_before}) and after (in \Cref{fig: chord_after}) the Merge in a split chord. In the bipartite chord diagrams, senders and receivers in OFAC-sanctioned transactions are arranged radially as arcs or nodes on the left and right sides between which the transactions are visualized by chords or links that connect them. Both before and after the Merge, the majority of OFAC-sanctioned transactions have a receiver with the sanctioned address with the last three digits of $f31$. Furthermore, \Cref{tab7} in \Cref{sec: sanctioned transactions} shows 9 transactions with unobserved delay for sanctioned transactions after the Merge, all associated with the sanctioned address with the last three digits of $31b$. No sanctioned transactions with unobserved delays before the Merge are found. Why? Figure 2 in ~\citet{yang2022sok} shows Flashbots as the major auction platform before the merge for private transactions. Flashbots is OFAC-compliant. In contrast, Table 7 in \citet{yang2022sok} shows the appearance of more censorship-free relays after the Merge, including bloXroute (Max profit and Ethical), Manifold, and Relayooor, which provides opportunities for OFAC-sanctioned transactions to proceed through private channels. 

\subsubsection{Implications}
\label{sec:confounding_implications}
Our results indicate that other confounding factors, such as OFAC sanction and private channel transactions, do not provide a significant or clear influence on waiting time so far. Compared with EIP upgrades and NFT drops, those factors are secondary to be considered in the design of future TFM. However, the OFAC sanction challenges the promise of blockchain governance to be censorship resistance~\cite{liu2022systematic,nabilou2020bitcoin}. We still need to watch out for future sanctions that might affect blockchain transactions significantly. On the other hand, the transaction waiting time via private channels, though currently unobservable, is an important tradeoff when choosing between private and public channels. Future research shall explore the method to study how the changes in private channel platforms affect the transaction latency on blockchains. 

\section{Design Block Interval for Transaction Fee Mechanism}
\label{sec: model}

\textbf{Ad Infinitum}: Can we abstract a mathematical model that explains the major impact of the Merge on TFM in particular and guides future TFM Design in general? 

\subsection{Three Major Protocol Changes of the Merge}
Since the transaction arrival rate---the average number of new transactions submitted to the public mempool per second---increases from 12.079 to 12.997\footnote{The numbers are calculated from the Ethereum mempool data. \citet{kapengut2023event} show that even the recorded transaction-per-day on Ethereum increases significantly after the Merge using simple statistics and t-test.}, the results must come from the protocol side. What are the specific protocol changes of the Merge that cause the reduction in high transaction waiting time, network loads, and market congestion? The Merge~\cite{merge2023} includes three major protocol changes: 
\begin{itemize}[leftmargin=*]
    \item {\em The consensus algorithm.} Ethereum changes the consensus algorithm from proof-of-work~\cite{wood2014ethereum} to a nested protocol that combines proof-of-stake and a finality gadget~\cite{buterin2020combining}. This also changes how miners are rewarded.
    \item {\em The block-building process.} The post-merge block building~\cite{blockbuild2023} allows validators to outsource block building to a network of builders via the builder market called MEV-Boost~\cite{MevBoost2023}. The builders produce entire blocks rather than just bundles of transactions---that way, builders could prevent risking the MEV from being stolen by producing a block header and getting this signed by the proposer without first releasing the body of the block. Notably, validators in general, do not get blocks directly from builders but mediated through MEV replays, the trusted parties who compete to be trustworthy.  
    \item {\em The block time or interval.} the Merge also changes the random block times, a Poisson process with an expectation of roughly 14 seconds, to fixed block times occurring every 12 seconds---or in consensus layer lingo coined as "slots." \cite{BlockTime2013} shows that less than 1\% blocks have a block interval of 24 or 36 seconds due to empty slots. The majority of blocks have a fixed block interval of 12 seconds.    
\end{itemize}
\vspace{3mm}
\subsection{The Mathematical Model}
Which of the three protocol changes lead to our results? We reason that our results are most likely to be the direct consequence of the block interval change. \citet{alharby2023transaction} systematically analyzes the factors that affect transaction latency, interchangeable with transaction waiting time in this study.\footnote{The definition of transaction latency or waiting time in \citet{alharby2023transaction} slightly differs from ours by also including the duration of transaction propagation, often called network latency, as shown in \Cref{figWT}.} In a simulation experiment based on queuing model, \citet{alharby2023transaction} evidence that transaction fees, block limit, block interval, transaction arrival rate (measured by TPS), and the behavior of the network nodes all contribute to transaction latency. However, \citet{alharby2023transaction} is not based on the EIP-1559 TFM. We elaborate that the base fee adjustment in EIP1559 TFM is crucial for the Merge, precisely the block interval reduction, to positively affect waiting time, network loads, and market congestion simultaneously. 

\begin{table}
  \caption{Notations}
    \centering
    \begin{tabular}{|p{2cm}||p{5cm}|}
    \hline
        Notation & Definition  \\ \hline
        $b$ & the base fee in EIP1559 TFM \\ \hline
       $D(b)$ & the total demand or transaction arrival rate \textbf{per second} measured in gas used that has a value, willingness to pay, or bid higher than the base fee $b$ \\\hline
       $I$ & the block interval\\\hline
    \end{tabular}
    \label{tab25}
\end{table}

\Cref{tab25} presents the notations.\footnote{$D(b)$ is monotonically decreasing in $b$.} The EIP-1559 Transaction Fee Mechanism is updated by block following the adjustment function below: 
\begin{equation}
      \text{b}_{n+1} = \text{b}_n (1+ \frac{1}{8} \frac{\text{GasUsed}_n - \text{GasTarget}}{\text{GasTarget}}).
      \label{eq:base_fee_adjustment}
\end{equation}
In the stable equilibrium where the base fee stop adjusting. We have:
\begin{equation}
    D^{e}(b^{e})\times I = \text{GasUsed}_n = \text{GasTarget}, 
    \forall n
\end{equation}
Thus,
\begin{equation}
    D^{e}(b^{e})\times I = \text{GasTarget}, 
    \forall n
\end{equation}
\subsubsection{Insight 1: Block Interval Changes}
 For protocol changes like the merge that reduces the block interval, EIP1559 TFM advises the base fee to decrease in adjusting to the new equilibrium. Future EIP proposals could accelerate this process by simultaneously decreasing the block interval and decreasing the base fee. 
 For instance, suppose that the block interval $I$ $\downarrow$ decreases to $I^{'}$ at block number $n$, i.e. $I^{'}< I$, when $b_{n}=b^{e}$, we have 
 
 \begin{equation}
    \text{GasUsed}_{n}=D^{e}(b^{e})\times I^{'} < D^{e}(b)\times I = \text{GasTarget}
    \label{eq:1}
 \end{equation}
Then \Cref{eq:base_fee_adjustment} and \Cref{eq:1} imply that:

 \begin{equation}
    \text{b}_{n+1} < \text{b}^{e}
 \end{equation}
We can accelerate the process by directly decreasing the base fee $b^{e}$ $\downarrow$ to $b^{e'}$, which increases the demand from $D^{e}(b^{e})$ $\uparrow$ to $D^{e}(b^{e'})$ such that 
 \begin{equation}
    \text{GasUsed}_{n}=D^{e}(b^{e'})\times I^{'} = D^{e}(b^{e})\times I = \text{GasTarget}
 \end{equation}

 \subsubsection{Insight 2: Demand Surges}
 For demand surges due to community events, EIP1559 TFM advises the base fee to increase in adjusting to the new equilibrium. Future EIP proposals could accelerate this process by increasing the base fee more frequently, in larger increments, or even ex-ante if foreseeing community events that create demand surges such as the NFT drops documented by our empirical results.
 
 For instance, when $D^{e}(\cdot)$ $\uparrow$ increases drastically to $D^{'}(\cdot)$ at block number $n$ at any base fee level, i.e. $D^{'}(b)\leq D^{e}(b)$ $\forall$ $b$ and $D^{'}(b)> D^{e}(b)$ for some $b$, when $b_{n}=b^{e}$, we have:
 
 \begin{equation}
    \text{GasUsed}_{n}=D^{'}(b^{e})\times I> D^{e}(b^{e})\times I = \text{GasTarget}
    \label{eq:2}
 \end{equation}
 
 Then \Cref{eq:base_fee_adjustment} and \Cref{eq:2} imply that:
 
  \begin{equation}
\text{b}_{n+1} > \text{b}^{e} 
 \end{equation}
We could accelerate the process by directly increasing the base fee from $b^{e}$ $\uparrow$ to $b^{e'}$, which coordinates the excess demand from $D^{e}(b^{e})$ $\downarrow$ to $D^{'}(b^{e'})$ such that:
 \begin{equation}
    \text{GasUsed}_{n}=D^{'}(b^{e'})\times I= D^{e}(b^{e})\times I = \text{GasTarget}
    \label{eq:3}
 \end{equation}

 \subsubsection{Insight 3: \\ How Block Interval Changes Affect Base Fee Adjustments During Demand Surges} 
\begin{figure}[!htbp]
	\includegraphics[width = \linewidth]{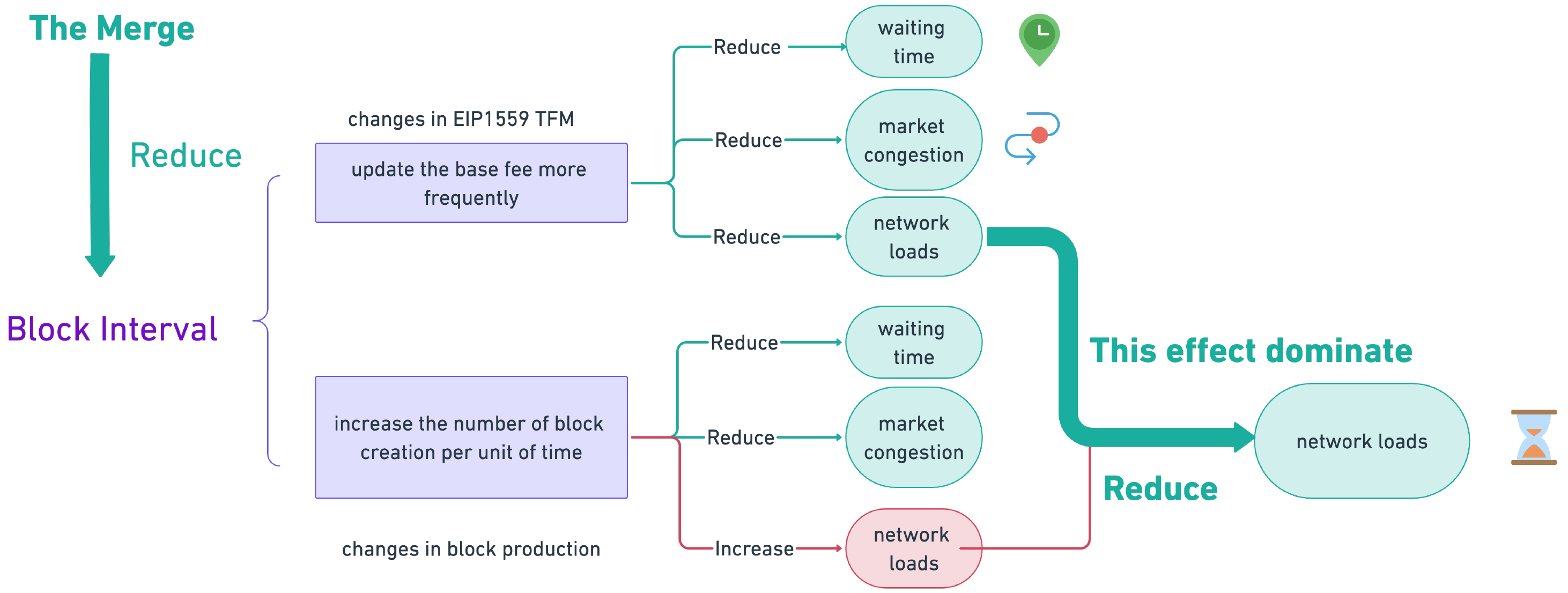}
	\caption{The Causal Graph for the Effect of Block Interval Reduction of the Merge}
	\label{causal_graph}
\end{figure}

For protocol changes like the merge that reduces the block interval, EIP1559 TFM would then adjusts the base fee much more frequently to the stable equilibrium during demand surges, which would reduce the market congestion, transaction waiting time, and network loads during the adjustment period as documented in our empirical results. We demonstrate the causal graph in \Cref{causal_graph}. In the case of demand surges, as shown in \Cref{eq:3}, the new stable equilibrium price $b^{e'}>b^{e}$, where $D^{'}(b^{e'})\times I= D^{e}(b^{e})\times I$ is independent of the block interval $I$. By \Cref{eq:base_fee_adjustment} and \Cref{eq:2}, we have
 \begin{equation}
 \begin{aligned}
    \frac{(\text{b}_{n+1}-\text{b}_n)}{\text{b}_n}
      = \frac{1}{8} \frac{\text{GasUsed}_n - \text{GasTarget}}{\text{GasTarget}}\\
       = \frac{1}{8} \frac{(D^{'}(b^{e})\times I-D^{e}(b^{e})\times I)}{D^{e}(b^{e})\times I}\\
       =\frac{1}{8} \frac{(D^{'}(b^{e})-D^{e}(b^{e}))}{D^{e}(b^{e})}.
\end{aligned}
\label{eq:4}
\end{equation}
 The percentages of base fee adjustment from block $n$ to $n+1$ as shown in \Cref{eq:4} are independent of blockchain interval. However, the adjustment happens more frequently with smaller block intervals. Thus, the base fee adjusts much faster after the Merge than before the merge, better signifying the high opportunity cost of time to transact on the Ethereum blockchain at this moment, which rejected recording the transaction arrivals from users who have values of transactions lower than the current opportunity cost.

\begin{figure*}[!htbp]
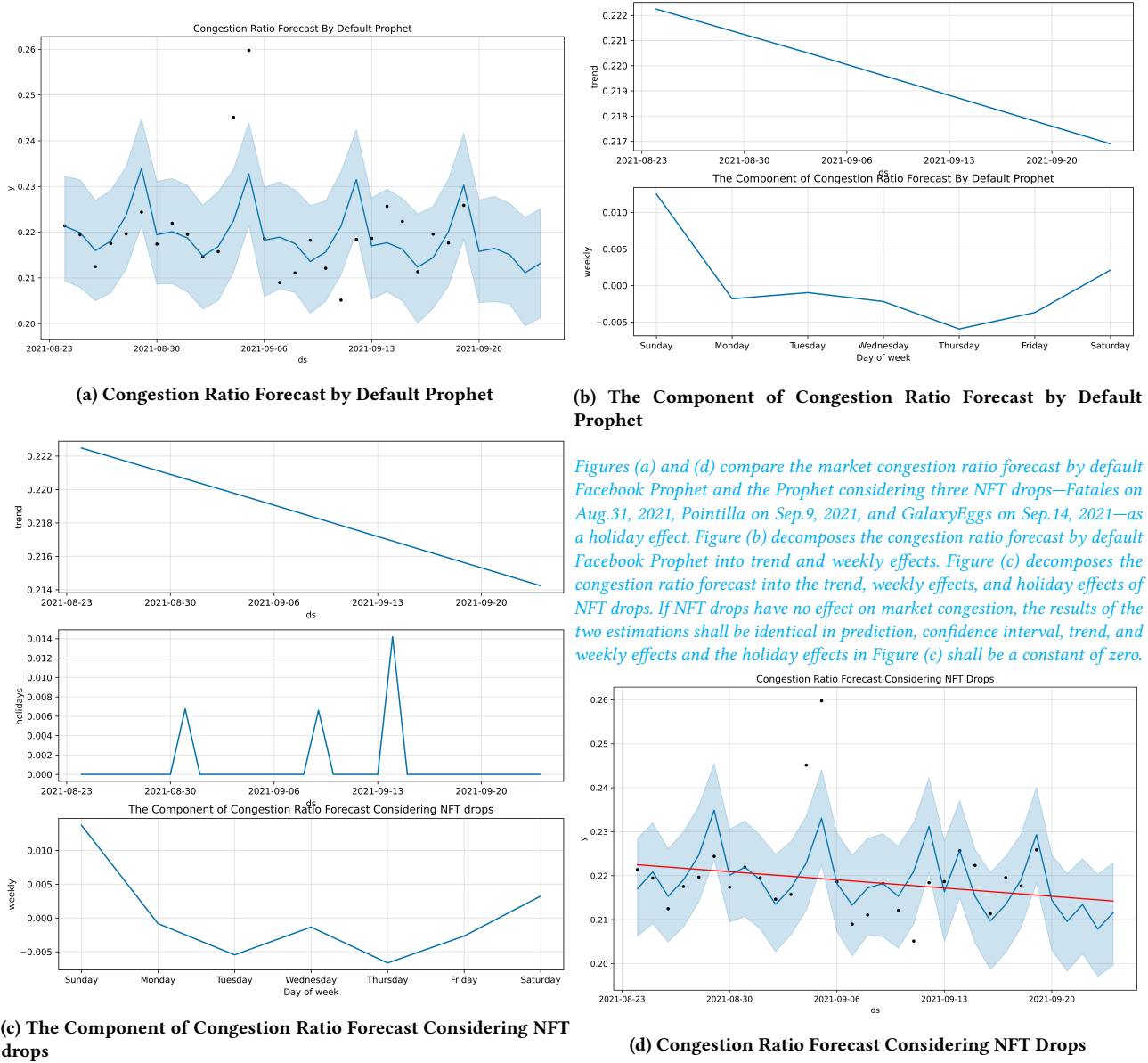

\begin{subfigure}{\columnwidth}
\includegraphics[width = \linewidth]{figs/figP1.pdf}
\caption{Congestion Ratio Forecast by Default Prophet}
\label{fig25}
 \end{subfigure}
\begin{subfigure}{\columnwidth}
\includegraphics[width = \linewidth]{figs/figP2.pdf}
\caption{The Component of Congestion Ratio Forecast by Default Prophet}
\label{fig26}
\end{subfigure}
\begin{subfigure}{\columnwidth}
\includegraphics[width = \linewidth]{figs/figP3.pdf}
\caption{The Component of Congestion Ratio Forecast Considering NFT drops}
\label{fig27}
\end{subfigure}
 \begin{subfigure}{\columnwidth}
\subcaption*{\small \textmd{\textit{\color{cyan}{Figures (a) and (d) compare the market congestion ratio forecast by default Facebook Prophet and the Prophet considering three NFT drops---Fatales on Aug.31, 2021, Pointilla on Sep.9, 2021, and GalaxyEggs on Sep.14, 2021---as a holiday effect. Figure (b) decomposes the congestion ratio forecast by default Facebook Prophet into trend and weekly effects. Figure (c) decomposes the congestion ratio forecast into the trend, weekly effects, and holiday effects of NFT drops. If NFT drops have no effect on market congestion, the results of the two estimations shall be identical in prediction, confidence interval, trend, and weekly effects and the holiday effects in Figure (c) shall be a constant of zero.}}}}
\includegraphics[width = \linewidth]{figs/figP4.pdf}
\caption{Congestion Ratio Forecast Considering NFT Drops}
\label{fig28}
\end{subfigure}
\caption{Congestion Ratio Forecast}
\label{fig:Congestion Ratio Forecast}
\end{figure*}

The design of time intervals between successive auctions to achieve the desired outcome has been discussed in the literature of frequent batch auctions (FBA)~\cite{budish2014implementation, budish2015high,jagannathan2022frequent}~\footnote{In the Ethereum ecosystem, the Cow protocol uses FBA as a solution to the issues of Miner Extractable Value (MEV): \url{https://docs.cow.fi/overview/batch-auctions}.} for the market design issues regarding high-frequency tradings. However, our analysis differs essentially in the two facets of auction format and design objectives: First, \citet{budish2014implementation} defines FBA as uniform-price sealed-bid double auctions conducted at frequent but discrete time intervals. In contrast, ours apply to the EIP1559 TFM---conducted at discrete time intervals---is neither uniform-price nor sealed-bid double auctions. Second, FBA was proposed to eliminate the speed race and the associated harm to liquidity and social welfare caused by continuous limit order books. Both \citet{budish2014implementation} and \citet{jagannathan2022frequent} suggest a longer block interval for preventing price manipulation during the time interval. In contrast, ours suggest a shorter time interval in EIP1559 TFM for more agile price adjustments during demand surges to reduce market congestion. 


\section{NFT Drops as A Case of Demand Surge}
\label{sec: NFT}
\textbf{Per Mutatis Mutandis}: how do NFT drops interact with market congestion in the EIP-1559 transaction fee mechanism?
\subsection{Results}
\label{sec: NFT_results}

\begin{figure*}[!htbp]
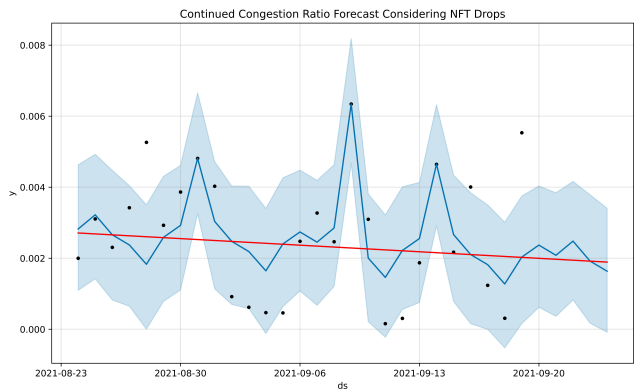

\begin{subfigure}{\columnwidth}
	\includegraphics[width = \linewidth]{figs/figP5.pdf}
	\caption{Continued Congestion Ratio Forecast by Default Prophet}
	\label{fig29}
\end{subfigure}
\begin{subfigure}{\columnwidth}
	\includegraphics[width = \linewidth]{figs/figP6.pdf}
	\caption{The Component of Continued Congestion Ratio Forecast by Default Prophet}
	\label{fig30}
\end{subfigure}
\begin{subfigure}{\columnwidth}
	\includegraphics[width = \linewidth]{figs/figP7.pdf}
	\caption{The Component of Continued Congestion Ratio Forecast Considering NFT drops}
	\label{fig31}
\end{subfigure}
\begin{subfigure}{\columnwidth}
\subcaption*{\small \textmd{\textit{\color{cyan}{Figures (a) and (d) compare the continued market congestion ratio forecast by default Facebook Prophet and the Prophet considering three NFT drops---Fatales on Aug.31, 2021, Pointilla on Sep.9, 2021, and GalaxyEggs on Sep.14, 2021---as a holiday effect. Figure (b) decomposes the continued congestion ratio forecast by default Facebook Prophet into trend and weekly effects. Figure (c) decomposes the continued congestion ratio forecast into the trend, weekly effects, and holiday effects of NFT drops. If NFT drops have no effect on market congestion, the results of the two estimations shall be identical in prediction, confidence interval, trend, and weekly effects, and the holiday effect in Figure (c) shall be a constant of zero.}}}}
	\includegraphics[width = \linewidth]{figs/figP8.pdf}
	\caption{Continued Congestion Ratio Forecast Considering NFT Drops}
	\label{fig32}
\end{subfigure}
\caption{Continued Congestion Ratio Forecast}
\label{fig:Continued Congestion Ratio Forecast}
\end{figure*}

We define the congestion ratio as the intraday percentage of congested blocks. And we define the continued congestion ratio as the intraday percentage of blocks being continued congested for 5 blocks. \Cref{fig:Congestion Ratio Forecast} compares the congestion ratio and its component forecast by default prophet and by the time-series model of holiday effects considering NFT drops: Fatales on Aug.31, 2021, Pointilla on Sep.9, 2021, and GalaxyEggs on Sep.14, 2021. If NFT drops have no effect on market congestion, the results of the two estimations shall be identical in prediction, confidence interval, trend, and weekly effects and the holiday effect in \Cref{fig:Continued Congestion Ratio Forecast} (c) shall be a constant of zero. However, the component of the congestion ratio forecast demonstrates obvious positive holiday effects on the dates of three NFT drops. Moreover, the NFT drops successfully match the peaks in congestion ratios.
\Cref{fig:Continued Congestion Ratio Forecast} compares the continued congestion ratio and its component forecast by default prophet and by the time-series model of holiday effect considering NFT drops. The time-series model again successfully identifies the three NFT drops as holiday effects besides the trend and seasonal effects. Moreover, the NFT drops successfully match the peaks in continued congestion ratios.

\subsection{Implications}
\label{sec: NFT_implications}
Our results show that NFT drops are a significant source of market congestion. Our study thus joins the existing literature in addressing the importance of NFT on blockchain efficiency and security. The NFT ecosystem is mushrooming, of which the market value excelled by 2 billion US dollars by 2021\cite{nadini2021mapping,wang2021non}. \citet{das2021understanding} first study the security issues of the NFT ecosystem. After examining the top 8 NFT marketplaces, \citet{das2021understanding} identify severe fraudulent user behaviors including counterfeit NFT creation and trading malpractices (e.g. wash trading, shill bidding, and bid shielding). Our result implies that the fraudulent user behaviors evidenced in \citet{das2021understanding} not only  contaminate blockchain integrity or security directly but could further worsen blockchain security and efficiency by creating market congestion.


\section{The Future of Transaction Fee Mechanism Design}
\label{sec:future TFM}
 
Our findings shed light on a few new directions for TFM design. Our study identifies two types of events that could significantly affect the transaction waiting time and other intertwined factors such as network loads and market congestion: EIP deployments and community-wide events such as NFT drops. Most importantly, we found the Merge reduces the risks of long waiting time, network loads, and market congestion on Ethereum. After examining three major protocol changes during the merge, we identify block interval shortening as the most plausible cause for our empirical results. Furthermore, in a mathematical model, we show changing block interval as a unique mechanism design choice for EIP1559 TFM. Reducing the block interval accelerates the base fee adjustments in EIP1559 TFM and coordinates the fee markets more promptly to achieve equilibrium, by which stakeholders enjoy improvements in both efficiency and security for trading on Ethereum. Our theoretical insights are generally applicable to any kind of demand surge. How frequent are demand spikes in Ethereum transactions? Furthermore, we have identified NFT drops as a unique source of market congestion beyond trend and season effects. To design the block interval for EIP1559 TFM could generally mitigate the market congestion caused by community-wide events including NFT drops. How can we further improve the protocol design to better coordinate the transaction fee market during demand surges ex-ante? Below, we elaborate on three future research directions.


\subsection{Causal inference study of EIP deployment}
Future research can extend our methodology to study other   EIP upgrades. For instance, articles in \cite{MergeCompete} demonstrate a variety of other changes during the Merge, including the increase in block building outsourcing, the increase in block inclusion time, the decrease in execution reward, and so forth by presenting visualizations and simple statistics, which can be further extended to draw causal inferences. Future research could also apply methods in causal machine learning such as representation learning~\cite{bengio2013representation} to illicit causal factors from various less explainable factors.    

\subsection{Community-wide events and TFMs}
Community-wide events such as popular NFT drops can cause market congestion. 
On the one hand, as shown in \citet{das2021understanding}, part of the demand surges during NFT drops are caused by scalping bots, which can be potentially removed by employing a Sybil-resistant decentralized identity protocol~\cite{maram2021candid}.~\footnote{https://nft.candid.id/}
On the one hand, the demand surges during NFT drops are analogous to the traffic surges during rush hours. Exploring the application of congestion pricing~\cite{hall2018pareto, mayer2003network, anderson2014subways} in market design in TFM is therefore an interesting future direction.

\subsection{Incorporating off-chain data in TFMs}
Although events such as NFT drops are usually scheduled in advance, existing TFMs do not make use of such information to adapt. Combining the ability to access off-chain data through oracles~\cite{breidenbach2021chainlink,DBLP:conf/ccs/ZhangMMGJ20,DBLP:conf/ccs/ZhangCCJS16} and on-chain mechanisms to design dynamic policies (e.g., TFMs) that can proactively respond to off-chain events is interesting future work.

\section*{Acknowledgments}
We thank Barnabé Monnot for his invaluable support and insightful comments throughout the research process. This research is an independent project supported by Ethereum Academic Grants Round.

\balance
\bibliographystyle{ACM-Reference-Format}
\bibliography{cite}

\newpage
\appendix
\onecolumn
\section{Glossary Table}
\label{appendix:glossary table}
\begin{table}[!htbp]
\fontsize{7pt}{\baselineskip}\selectfont
\setlength{\tabcolsep}{1pt}
\begin{tabular}{|p{2cm}||p{10cm}|p{2cm}|p{2cm}|}
 \hline \hline
Terms & Definition & Source & Field \\ [1ex] 
 \hline \hline
 \multicolumn{3}{c}{\textbf{Term 1: Latency}} \\ [1ex] 
 \hline
direct latency & the latency with which messages reach a listener node from one or more vantage points—in other words, source latency. & \citet{tang_2022_strategic} & distributed system\\[0.2ex] 

transaction latency  & the duration from transaction creation to transaction inclusion in the confirmed block & \citet{alharby2023transaction} & distributed system\\[0.2ex] 

Decision Latency & the latency that corresponds to a node’s ability to inject itself between a pair of communicating nodes. The starting point is when a High-frequency Trading (HFT) firm’s resting limit order is executed by an incoming market order. The matching engine processes and time stamps the trade. A confirmation message is then sent to the HFT firm. The firm processes the confirmation information and makes a decision on how to react, which may be in the form of an aggressive order. The end of the latency measure is marked by the time stamp assigned when the message for the market order is processed by the matching engine.. & \citet{baron2019risk} & HFT, Finance\\[0.2ex] 

 \hline
\multicolumn{3}{c}{\textbf{Term 2: Waiting Time}} \\ [1ex] 
 \hline
waiting time & the duration between when a service is requested and when a service is performed & \citet{baker1996effects} & marketing\\[0.2ex] 
response time & response time is measured as the length of time from the moment a problem is sent to a subject until his response is received by the server & \citet{rubinstein2013response} & decision theory\\[0.2ex] 
block interval/time & the time between consecutive blocks & \citet{Yaish2022EC} & distributed system\\[0.2ex]

\hline
\multicolumn{3}{c}{\textbf{Term 3: Market Congestion}} \\ [1ex] 
 \hline 
congested state & overloading and voltage violations in a distributed network & \citet{attar2022congestion} & Distributed System \\[0.2ex] 

market congestion  & the time that the market clears, between the time that market participants enter and the time that exchanges are settled. & \citet{roth2008have} & Market Design \\[0.2ex] 

traffic congestion  & the time difference between the desired travel time and the actual travel time during rush hours & \citet{hall2018pareto}& Market Design \\[0.2ex] 

 \hline
\end{tabular}
\caption{The Glossary Table}
\label{tab18}
\end{table}

\newpage
\section{Data Dictionary}
\label{sec: data dictionary}
\begin{table}[!htbp]
\fontsize{7pt}{\baselineskip}\selectfont
\setlength{\tabcolsep}{1pt}
\begin{tabular}{|c|c|c|c|}
 \hline \hline
 Column Name & Source & Type & Annotation \\ [1ex] 
 \hline \hline
 \multicolumn{3}{c}{\textbf{Data Source 1: Ethereum Blockchain Data (blockchain)}} \\ [1ex] 
 \hline
number & Ethereum & integer & Block number on blockchain\\[0.2ex] 
gas\_limit & Ethereum & integer & The maximum gas allowed in this block\\[0.2ex] 
gas\_used & Ethereum & integer & The total used gas by all transactions in this block\\[0.2ex] 
transaction\_count & Ethereum & integer & Number of all transactions included in this block\\[0.2ex] 
timestamp & Ethereum & integer & Unix Timestamp\\[0.2ex] 
base\_fee\_per\_gas & Ethereum & integer & Protocol base fee per gas\\[0.2ex] 
 \hline
\multicolumn{3}{c}{\textbf{Data Source 1: Ethereum Blockchain Data (transaction)}} \\ [1ex] 
 \hline
block\_number & Ethereum & integer & Block number on blockchain\\[0.2ex] 
hash & Ethereum & varchar(80) & the transaction hash (key) on blockchain\\[0.2ex] 
from\_address & Ethereum & varchar(42) & the public address (key) of the transaction sender\\[0.2ex] 
to\_address & Ethereum & varchar(42) & the public address (key) of the transaction receiver\\[0.2ex] 
\hline
\multicolumn{3}{c}{\textbf{Data Source 2: MemopoolGuru}} \\ [1ex] 
 \hline 
included\_in\_block\_num & MemopoolGuru & integer & block number on blockchain where the transaction is included \\[0.2ex] 
delay  & MemopoolGuru & double precision & the transaction waiting time (delay) in second \\[0.2ex] 
hash  & MemopoolGuru & varchar(80) & the transaction hash (key) on blockchain \\[0.2ex] 
 \hline
\multicolumn{3}{c}{\textbf{Data Source 3:  OFAC Sanction Program}} \\
 \hline
   address & OFAC  & varchar(42) & the public address (key) of the sanctioned address\\[0.2ex] 
\hline
\end{tabular}
\caption{Data dictionary}
\label{tab12}
\end{table}
\newpage

\section{Statistics}
\label{sec:statistics}
\begin{table}[!htbp] \centering
  \caption{Statistics for Gas Used Per Second by Block Before and After the Merge}
    \begin{tabular}{rrrrrr}
    \toprule
    Gas Used Per Second &  median &  25\% quantile & 75\% quantile &  mean &  std \\
    \midrule
      before the Merge &                1.111929e+06 &             761741.537500 &              1.851871e+06 &              1.815617e+06 &             2.523168e+06 \\
      after the Merge &                1.204980e+06 &             730119.229167 &              1.805011e+06 &              1.270889e+06 &             7.034426e+05 \\

    \bottomrule
    \end{tabular}
\label{tab8}
\end{table}
\begin{table}[!htbp] \centering
  \caption{Statistics for The Moving Average (5 Blocks) of Gas Used Per Second by Block Before and After the Merge}
    \begin{tabular}{rrrrrr}
    \toprule
     Gas Used Per Second MA5 &  median &  25\% quantile &  75\% quantile &  mean &  std \\
    \midrule
          before the Merge &                 1.493161e+06 &               1.152283e+06 &               2.081531e+06 &               1.815480e+06 &              1.083907e+06 \\
          after the Merge &                 1.243960e+06 &               1.108872e+06 &               1.396571e+06 &               1.271037e+06 &              2.493085e+05 \\
    \bottomrule
    \end{tabular}
\label{tab9}
\end{table}
\begin{table}[!htbp] \centering
  \caption{Statistics for The Moving Average (7200 Blocks) of Gas Used Per Second by Block Before and After the Merge}
    \begin{tabular}{rrrrrr}
    \toprule
      Gas Used Per Second MA7200 &  median &  25\% quantile &  75\% quantile &  mean &  std \\
    \midrule
          before the Merge &                    1.493161e+06 &                  1.152283e+06 &                  2.081531e+06 &                  1.815480e+06 &                 1.083907e+06 \\
          after the Merge &                    1.243960e+06 &                  1.108872e+06 &                  1.396571e+06 &                  1.271037e+06 &                 2.493085e+05 \\
    \bottomrule
    \end{tabular}
\label{tab20}
\end{table}

\newpage
\section{Regression Tables}
\label{sec: regression tables}
\begin{table}[!htbp] \centering
  \caption{The Logit Regression for Market Congestion (>95\% gas used) before and after the merge}
\begin{tabular}{@{\extracolsep{5pt}}lccc}
\\[-1.8ex]\hline
\hline \\[-1.8ex]
& \multicolumn{3}{c}{\textit{Dependent variable:}} \
\cr \cline{3-4}
\\[-1.8ex] & (1) & (2) & (3) \\
\hline \\[-1.8ex]
 merged & -0.934$^{***}$ & -0.778$^{***}$ & -0.749$^{***}$ \\
  & (0.016) & (0.030) & (0.031) \\
 blockn & & -0.000$^{***}$ & -0.000$^{***}$ \\
  & & (0.000) & (0.000) \\
 merged:blockn & & & -0.000$^{***}$ \\
  & & & (0.000) \\
 Intercept & -1.226$^{***}$ & -1.304$^{***}$ & -1.268$^{***}$ \\
  & (0.009) & (0.016) & (0.019) \\
\hline \\[-1.8ex]
 Observations & 134,831 & 134,831 & 134,831 \\
 $R^2$ &  &  &  \\
 Adjusted $R^2$ &  &  &  \\
 Residual Std. Error & 1.000(df = 134829) & 1.000(df = 134828) & 1.000(df = 134827)  \\
 F Statistic & $^{}$ (df = 1.0; 134829.0) & $^{}$ (df = 2.0; 134828.0) & $^{}$ (df = 3.0; 134827.0) \\
\hline
\hline \\[-1.8ex]
\textit{Note:} & \multicolumn{3}{r}{$^{*}$p$<$0.1; $^{**}$p$<$0.05; $^{***}$p$<$0.01} \\
\end{tabular}
\label{tab1}
\end{table}
\begin{table}[!htbp] \centering
  \caption{The Logit Regression for Market Continued Congestion (>95\% gas used) before and after the merge}
\begin{tabular}{@{\extracolsep{5pt}}lccc}
\\[-1.8ex]\hline
\hline \\[-1.8ex]
& \multicolumn{3}{c}{\textit{Dependent variable:}} \
\cr \cline{3-4}
\\[-1.8ex] & (1) & (2) & (3) \\
\hline \\[-1.8ex]
 merged & -0.633$^{***}$ & -0.556$^{***}$ & -0.529$^{***}$ \\
  & (0.083) & (0.160) & (0.164) \\
 blockn & & -0.000$^{}$ & -0.000$^{}$ \\
  & & (0.000) & (0.000) \\
 merged:blockn & & & -0.000$^{}$ \\
  & & & (0.000) \\
 Intercept & -5.078$^{***}$ & -5.117$^{***}$ & -5.083$^{***}$ \\
  & (0.049) & (0.085) & (0.098) \\
\hline \\[-1.8ex]
 Observations & 134,831 & 134,831 & 134,831 \\
 $R^2$ &  &  &  \\
 Adjusted $R^2$ &  &  &  \\
 Residual Std. Error & 1.000(df = 134829) & 1.000(df = 134828) & 1.000(df = 134827)  \\
 F Statistic & $^{}$ (df = 1.0; 134829.0) & $^{}$ (df = 2.0; 134828.0) & $^{}$ (df = 3.0; 134827.0) \\
\hline
\hline \\[-1.8ex]
\textit{Note:} & \multicolumn{3}{r}{$^{*}$p$<$0.1; $^{**}$p$<$0.05; $^{***}$p$<$0.01} \\
\end{tabular}
\label{tab2}
\end{table}
\begin{table}[!htbp] \centering
  \caption{The 75\% Quantile of Waiting Time (delay) before and after the merge}
\begin{tabular}{@{\extracolsep{5pt}}lccc}
\\[-1.8ex]\hline
\hline \\[-1.8ex]
& \multicolumn{3}{c}{\textit{Dependent variable:}} \
\cr \cline{3-4}
\\[-1.8ex] & (1) & (2) & (3) \\
\hline \\[-1.8ex]
 merged & -13.844$^{***}$ & -11.975$^{***}$ & -13.421$^{***}$ \\
  & (0.048) & (0.098) & (0.098) \\
 blockn & & -0.000$^{***}$ & 0.000$^{***}$ \\
  & & (0.000) & (0.000) \\
 merged:blockn & & & -0.000$^{***}$ \\
  & & & (0.000) \\
 Intercept & 34.365$^{***}$ & 33.423$^{***}$ & 35.039$^{***}$ \\
  & (0.034) & (0.055) & (0.069) \\
\hline \\[-1.8ex]
 Observations & 134,831 & 134,831 & 134,831 \\
 $R^2$ & nan & nan & nan \\
 Adjusted $R^2$ & nan & nan & nan \\
 Residual Std. Error & 1.000(df = 134829) & 1.000(df = 134828) & 1.000(df = 134827)  \\
 F Statistic & nan$^{***}$ (df = 1.0; 134829.0) & nan$^{***}$ (df = 2.0; 134828.0) & nan$^{***}$ (df = 3.0; 134827.0) \\
\hline
\hline \\[-1.8ex]
\textit{Note:} & \multicolumn{3}{r}{$^{*}$p$<$0.1; $^{**}$p$<$0.05; $^{***}$p$<$0.01} \\
\end{tabular}
\label{tab4}
\end{table}

\begin{table}[!htbp] \centering
  \caption{IQR of Waiting Time (delay) before and after the merge}
\begin{tabular}{@{\extracolsep{5pt}}lccc}
\\[-1.8ex]\hline
\hline \\[-1.8ex]
& \multicolumn{3}{c}{\textit{Dependent variable:}} \
\cr \cline{3-4}
\\[-1.8ex] & (1) & (2) & (3) \\
\hline \\[-1.8ex]
 merged & -32.269$^{***}$ & -26.011$^{***}$ & -26.057$^{***}$ \\
  & (1.140) & (2.276) & (2.276) \\
 blockn & & -0.000$^{***}$ & 0.000$^{}$ \\
  & & (0.000) & (0.000) \\
 merged:blockn & & & -0.000$^{***}$ \\
  & & & (0.000) \\
 Intercept & 52.611$^{***}$ & 49.487$^{***}$ & 54.509$^{***}$ \\
  & (0.807) & (1.272) & (1.612) \\
\hline \\[-1.8ex]
 Observations & 134,831 & 134,831 & 134,831 \\
 $R^2$ & 0.006 & 0.006 & 0.006 \\
 Adjusted $R^2$ & 0.006 & 0.006 & 0.006 \\
 Residual Std. Error & 209.219(df = 134829) & 209.212(df = 134828) & 209.193(df = 134827)  \\
 F Statistic & 801.865$^{***}$ (df = 1.0; 134829.0) & 406.004$^{***}$ (df = 2.0; 134828.0) & 279.302$^{***}$ (df = 3.0; 134827.0) \\
\hline
\hline \\[-1.8ex]
\textit{Note:} & \multicolumn{3}{r}{$^{*}$p$<$0.1; $^{**}$p$<$0.05; $^{***}$p$<$0.01} \\
\end{tabular}
\label{tab5}
\end{table}
\begin{table}[!htbp] \centering
  \caption{The Median of Waiting Time (delay) before and after the merge}
\begin{tabular}{@{\extracolsep{5pt}}lccc}
\\[-1.8ex]\hline
\hline \\[-1.8ex]
& \multicolumn{3}{c}{\textit{Dependent variable:}} \
\cr \cline{3-4}
\\[-1.8ex] & (1) & (2) & (3) \\
\hline \\[-1.8ex]
 merged & -9.659$^{***}$ & -0.225$^{}$ & -0.239$^{}$ \\
  & (0.661) & (1.319) & (1.319) \\
 blockn & & -0.000$^{***}$ & -0.000$^{***}$ \\
  & & (0.000) & (0.000) \\
 merged:blockn & & & -0.000$^{***}$ \\
  & & & (0.000) \\
 Intercept & 28.142$^{***}$ & 23.434$^{***}$ & 25.016$^{***}$ \\
  & (0.468) & (0.737) & (0.934) \\
\hline \\[-1.8ex]
 Observations & 134,831 & 134,831 & 134,831 \\
 $R^2$ & 0.002 & 0.002 & 0.002 \\
 Adjusted $R^2$ & 0.002 & 0.002 & 0.002 \\
 Residual Std. Error & 121.289(df = 134829) & 121.258(df = 134828) & 121.255(df = 134827)  \\
 F Statistic & 213.775$^{***}$ (df = 1.0; 134829.0) & 141.072$^{***}$ (df = 2.0; 134828.0) & 96.588$^{***}$ (df = 3.0; 134827.0) \\
\hline
\hline \\[-1.8ex]
\textit{Note:} & \multicolumn{3}{r}{$^{*}$p$<$0.1; $^{**}$p$<$0.05; $^{***}$p$<$0.01} \\
\end{tabular}
\label{tab6}
\end{table}

\begin{table}[!htbp] \centering
  \caption{The gas used per second by block before and after the merge}
\begin{tabular}{@{\extracolsep{5pt}}lccc}
\\[-1.8ex]\hline
\hline \\[-1.8ex]
& \multicolumn{3}{c}{\textit{Dependent variable:}} \
\cr \cline{3-4}
\\[-1.8ex] & (1) & (2) & (3) \\
\hline \\[-1.8ex]
 merged & -544728.569$^{***}$ & -507576.370$^{***}$ & -507395.185$^{***}$ \\
  & (10074.034) & (20120.755) & (20120.609) \\
 blockn & & -0.531$^{**}$ & -1.097$^{***}$ \\
  & & (0.249) & (0.353) \\
 merged:blockn & & & 1.124$^{**}$ \\
  & & & (0.498) \\
 Intercept & 1815617.089$^{***}$ & 1797075.882$^{***}$ & 1777314.312$^{***}$ \\
  & (7135.295) & (11245.745) & (14247.892) \\
\hline \\[-1.8ex]
 Observations & 134,830 & 134,830 & 134,830 \\
 $R^2$ & 0.021 & 0.021 & 0.021 \\
 Adjusted $R^2$ & 0.021 & 0.021 & 0.021 \\
 Residual Std. Error & 1849542.236(df = 134828) & 1849517.888(df = 134827) & 1849489.752(df = 134826)  \\
 F Statistic & 2923.839$^{***}$ (df = 1.0; 134828.0) & 1464.233$^{***}$ (df = 2.0; 134827.0) & 977.886$^{***}$ (df = 3.0; 134826.0) \\
\hline
\hline \\[-1.8ex]
\textit{Note:} & \multicolumn{3}{r}{$^{*}$p$<$0.1; $^{**}$p$<$0.05; $^{***}$p$<$0.01} \\
\end{tabular}
\label{tab10}
\end{table}
\begin{table}[!htbp] \centering
  \caption{The moving average (5 blocks) of gas used per second before and after the merge}
\begin{tabular}{@{\extracolsep{5pt}}lccc}
\\[-1.8ex]\hline
\hline \\[-1.8ex]
& \multicolumn{3}{c}{\textit{Dependent variable:}} \
\cr \cline{3-4}
\\[-1.8ex] & (1) & (2) & (3) \\
\hline \\[-1.8ex]
 merged & -544442.231$^{***}$ & -506485.276$^{***}$ & -506303.778$^{***}$ \\
  & (4277.151) & (8542.016) & (8541.222) \\
 blockn & & -0.542$^{***}$ & -1.107$^{***}$ \\
  & & (0.106) & (0.150) \\
 merged:blockn & & & 1.123$^{***}$ \\
  & & & (0.211) \\
 Intercept & 1815479.678$^{***}$ & 1796536.988$^{***}$ & 1776803.682$^{***}$ \\
  & (3029.457) & (4774.227) & (6048.269) \\
\hline \\[-1.8ex]
 Observations & 134,831 & 134,831 & 134,831 \\
 $R^2$ & 0.107 & 0.107 & 0.108 \\
 Adjusted $R^2$ & 0.107 & 0.107 & 0.108 \\
 Residual Std. Error & 785266.518(df = 134829) & 785192.705(df = 134828) & 785113.422(df = 134827)  \\
 F Statistic & 16202.960$^{***}$ (df = 1.0; 134829.0) & 8116.178$^{***}$ (df = 2.0; 134828.0) & 5421.289$^{***}$ (df = 3.0; 134827.0) \\
\hline
\hline \\[-1.8ex]
\textit{Note:} & \multicolumn{3}{r}{$^{*}$p$<$0.1; $^{**}$p$<$0.05; $^{***}$p$<$0.01} \\
\end{tabular}
\label{tab11}
\end{table}

\begin{table}[!htbp] \centering
  \caption{The moving average (7200 blocks) of gas used per second before and after the merge}
\begin{tabular}{@{\extracolsep{5pt}}lccc}
\\[-1.8ex]\hline
\hline \\[-1.8ex]
& \multicolumn{3}{c}{\textit{Dependent variable:}} \
\cr \cline{3-4}
\\[-1.8ex] & (1) & (2) & (3) \\
\hline \\[-1.8ex]
 merged & -518322.895$^{***}$ & -390039.900$^{***}$ & -390194.973$^{***}$ \\
  & (412.804) & (718.969) & (711.180) \\
 blockn & & -1.833$^{***}$ & -1.350$^{***}$ \\
  & & (0.009) & (0.012) \\
 merged:blockn & & & -0.959$^{***}$ \\
  & & & (0.018) \\
 Intercept & 1818286.002$^{***}$ & 1754265.457$^{***}$ & 1771125.786$^{***}$ \\
  & (292.384) & (401.840) & (503.605) \\
\hline \\[-1.8ex]
 Observations & 134,831 & 134,831 & 134,831 \\
 $R^2$ & 0.921 & 0.940 & 0.941 \\
 Adjusted $R^2$ & 0.921 & 0.940 & 0.941 \\
 Residual Std. Error & 75788.939(df = 134829) & 66088.500(df = 134828) & 65371.995(df = 134827)  \\
 F Statistic & 1576572.834$^{***}$ (df = 1.0; 134829.0) & 1057921.096$^{***}$ (df = 2.0; 134828.0) & 721816.731$^{***}$ (df = 3.0; 134827.0) \\
\hline
\hline \\[-1.8ex]
\textit{Note:} & \multicolumn{3}{r}{$^{*}$p$<$0.1; $^{**}$p$<$0.05; $^{***}$p$<$0.01} \\
\end{tabular}
\label{tab22}
\end{table}

\newpage
\section{Sanctioned Transactions}
\label{sec: sanctioned transactions}
\begin{table}[!htbp]  \centering
  \caption{The Sanctioned Transaction with Unobserved Delay (only found after the merge): total number $=$9}
\begin{tabular}{ll}
\toprule
                                   address &                                                         hash \\
\midrule
0xd90e2f925da726b50c4ed8d0fb90ad053324f31b & 0x0fb9a4e23d7c504571daed1394feea8bcb823a18a2887e7cab54f7f... \\
0xd90e2f925da726b50c4ed8d0fb90ad053324f31b & 0xb37f8b00993271dd97ac9acdb4d84cb31cd210a037286bcc4bc7887... \\
0xd90e2f925da726b50c4ed8d0fb90ad053324f31b & 0xd5c80d358016a1fe77260fa9cab88a706337c6f41f9112eca8840e9... \\
0xd90e2f925da726b50c4ed8d0fb90ad053324f31b & 0x52cd1910b0710e73a9fb85f9e13861ef828a3244a1fa03944e47adf... \\
0xd90e2f925da726b50c4ed8d0fb90ad053324f31b & 0xe4a84c77748a627a093f79c6abb4b037c60ae9f58977bab2c46d17a... \\
0xd90e2f925da726b50c4ed8d0fb90ad053324f31b & 0x094cd7a9dc72cf6b3415fd18223a075e631d4ee5bfc900b9293dbd1... \\
0xd90e2f925da726b50c4ed8d0fb90ad053324f31b & 0x78c426d5b25290aa03153c728383f2d422c7a14a40a0a5e11bea67c... \\
0xd90e2f925da726b50c4ed8d0fb90ad053324f31b & 0x9d4fcf6315d8fbf2e07dc71a2c81df8c86a2dd518ac76c48eaeb397... \\
0xd90e2f925da726b50c4ed8d0fb90ad053324f31b & 0x0b9613f988da92901816bc76af743974032985b5eeeab70d5940c65... \\
\bottomrule
\end{tabular}
\label{tab7}
\end{table}
\begin{table}
\caption{Statistics of the Delay for Sanctioned and Unsanctioned Transactions Before and After the Merge}

\begin{tabular}{llrrrrrrrr}
\toprule
  &   &      count &         mean &            std &       min &        25\% &        50\% &        75\% &           max \\
sanctioned & merge &            &              &                &           &            &            &            &               \\
\midrule
no & before &  9712821.0 &   974.101368 &   37825.688389 &  0.000020 &   7.793996 &  15.191844 &  29.682377 &  5.441627e+06 \\
  & after &  9120233.0 &  4868.072870 &  152957.817814 &  0.000064 &   9.147091 &  13.792645 &  18.629303 &  6.292439e+06 \\
yes & before &     1007.0 &  6161.196202 &  131471.221952 &  0.115805 &  10.072948 &  19.687913 &  37.180833 &  2.950860e+06 \\
  & after &      466.0 &    47.262011 &     257.400994 &  1.036414 &  11.876102 &  17.153332 &  22.571568 &  3.107158e+03 \\
\bottomrule
\end{tabular}
    \label{tab17}
\end{table}

\end{document}